\documentclass[aps,prd,twocolumn,showpacs,groupeaddress,amsmath,amssymb]{revtex4}

\usepackage{graphicx}
\usepackage{dcolumn}
\usepackage{bm}
\usepackage{mathrsfs} 
\usepackage{xspace}
\usepackage{amssymb}
\usepackage{ifthen}

\newcommand{\etal}{{\sl et al.}}

\newcommand{\W}{\ensuremath{W}\xspace}

\newcommand{\psgn}{\ensuremath{P_{\rm sgn}}\xspace}
\newcommand{\pbkg}{\ensuremath{P_{\rm bkg}}\xspace}
\newcommand{\mtop}{\ensuremath{m_t}\xspace}

\newcommand{\ftop}{\ensuremath{f_{\rm top}}\xspace}

\newcommand{\qqbar}{\ensuremath{q\bar{q}}\xspace}
\newcommand{\ppbar}{\ensuremath{p\bar{p}}\xspace}

\newcommand{\ttbar}{\ensuremath{t\bar{t}}\xspace}

\newcommand{\wjets}{\ensuremath{W}+\rm jets\xspace}

\newcommand{\ljets}{\ensuremath{\ell}+\rm jets\xspace}
\newcommand{\etmiss}{\ensuremath{E \kern-0.6em\slash _T}\xspace}
\newcommand{\etmissx}{\ensuremath{E \kern-0.6em\slash_{\rm x}}\xspace}
\newcommand{\etmissy}{\ensuremath{E \kern-0.6em\slash_{\rm y}}\xspace}

\newcommand{\alpgen}{{\sc alpgen}\xspace}
\newcommand{\pythia}{{\sc pythia}\xspace}
\newcommand{\geant}{{\sc geant}\xspace}

\newcommand{\ipb}{\ensuremath{\rm pb^{-1}}\xspace}

\newcommand{\MeV}{\ensuremath{\mathrm{Me\kern-0.1em V}}\xspace}
\newcommand{\GeV}{\ensuremath{\mathrm{Ge\kern-0.1em V}}\xspace}
\newcommand{\GeVc}{\ensuremath{\mathrm{Ge\kern-0.1em V}}\xspace}
\newcommand{\GeVcc}{\ensuremath{\mathrm{Ge\kern-0.1em V}}\xspace}
\newcommand{\TeV}{\ensuremath{\mathrm{Te\kern-0.1em V}}\xspace}

\newcommand{\met}{\mbox{$\not\!\!E_T$}\ }

\newcommand{\MET}{$\not\!\!E_T$\ }

\newcommand{\jes}{{\it JES}}
 
\newcommand{\bi}{\begin{itemize}}
\newcommand{\ei}{\end{itemize}}

\begin{document}

% 2/7/07: This is now version 6.0
% THIS VERSION IS FOR FINAL 24h Collaboration review

% \leftline{Version 5.6 as of \today} 
% \leftline{Primary authors: P.~Houben, M.~Mulders, M.~Weber}
% \leftline{To be submitted to PRD}

% \rightline{Comment to {\tt d0-run2eb-011@fnal.gov}}
% \rightline{by Jan 24th, 2007}

% the following line is for submission
\hspace{5.2in} \mbox{FERMILAB-PUB-07-039-E}

\title{Measurement of the top quark mass in the lepton+jets channel using the Ideogram Method}

% throughout the internal review, the note will be authored by individuals
% who contributed to the paper
%
%\author{Pieter Houben}
%\affiliation{NIKHEF, phouben@fnal.gov}
%
%\author{Martijn Mulders}
%\affiliation{CERN, mulders@fnal.gov}
%
%\author{Michele Weber}
%\affiliation{FNAL, webermi@fnal.gov}
%\vspace*{0.5cm}

% once the note is approved for conferences, the following author will be used
% \author{The \dzero Collaboration}
% \affiliation{URL http://www-d0.fnal.gov}
% 2/7/07: removed for submission

% use the official authorlist for publication
% LIST_OF_AUTHORS_R2.TEX                2/6/07              
%
\author{                                                                      
%% names begin here                                                           
V.M.~Abazov,$^{35}$                                                           
B.~Abbott,$^{75}$                                                             
M.~Abolins,$^{65}$                                                            
B.S.~Acharya,$^{28}$                                                          
M.~Adams,$^{51}$                                                              
T.~Adams,$^{49}$                                                              
E.~Aguilo,$^{5}$                                                              
S.H.~Ahn,$^{30}$                                                              
M.~Ahsan,$^{59}$                                                              
G.D.~Alexeev,$^{35}$
% ...
\\                                                                            
\vskip 0.30cm                                                                 
\centerline{(D\O\ Collaboration)}                                             
\vskip 0.30cm                                                                 
}                                                                             
\affiliation{                                                                 
\centerline{$^{1}$Universidad de Buenos Aires, Buenos Aires, Argentina}       
\centerline{$^{2}$LAFEX, Centro Brasileiro de Pesquisas F{\'\i}sicas,         
                  Rio de Janeiro, Brazil}                                     
\centerline{$^{3}$Universidade do Estado do Rio de Janeiro,                   
                  Rio de Janeiro, Brazil}                                     
\centerline{$^{4}$Instituto de F\'{\i}sica Te\'orica, Universidade            
                  Estadual Paulista, S\~ao Paulo, Brazil}                     
\centerline{$^{5}$University of Alberta, Edmonton, Alberta, Canada,           
                  Simon Fraser University, Burnaby, British Columbia, Canada,}
\centerline{York University, Toronto, Ontario, Canada, and                    
                  McGill University, Montreal, Quebec, Canada}                
\centerline{$^{6}$University of Science and Technology of China, Hefei,       
                  People's Republic of China}                                 
\centerline{$^{7}$Universidad de los Andes, Bogot\'{a}, Colombia}             
\centerline{$^{8}$Center for Particle Physics, Charles University,            
                  Prague, Czech Republic}                                     
\centerline{$^{9}$Czech Technical University, Prague, Czech Republic}         
\centerline{$^{10}$Center for Particle Physics, Institute of Physics,         
                   Academy of Sciences of the Czech Republic,                 
                   Prague, Czech Republic}                                    
\centerline{$^{11}$Universidad San Francisco de Quito, Quito, Ecuador}        
\centerline{$^{12}$Laboratoire de Physique Corpusculaire, IN2P3-CNRS,         
                   Universit\'e Blaise Pascal, Clermont-Ferrand, France}      
\centerline{$^{13}$Laboratoire de Physique Subatomique et de Cosmologie,      
                   IN2P3-CNRS, Universite de Grenoble 1, Grenoble, France}    
\centerline{$^{14}$CPPM, IN2P3-CNRS, Universit\'e de la M\'editerran\'ee,     
                   Marseille, France}                                         
\centerline{$^{15}$Laboratoire de l'Acc\'el\'erateur Lin\'eaire,              
                   IN2P3-CNRS et Universit\'e Paris-Sud, Orsay, France}       
\centerline{$^{16}$LPNHE, IN2P3-CNRS, Universit\'es Paris VI and VII,         
                   Paris, France}                                             
\centerline{$^{17}$DAPNIA/Service de Physique des Particules, CEA, Saclay,    
                   France}                                                    
\centerline{$^{18}$IPHC, IN2P3-CNRS, Universit\'e Louis Pasteur, Strasbourg,  
                   France, and Universit\'e de Haute Alsace,                  
                   Mulhouse, France}                                          
\centerline{$^{19}$IPNL, Universit\'e Lyon 1, CNRS/IN2P3, Villeurbanne, France
                   and Universit\'e de Lyon, Lyon, France}                    
\centerline{$^{20}$III. Physikalisches Institut A, RWTH Aachen,               
                   Aachen, Germany}                                           
\centerline{$^{21}$Physikalisches Institut, Universit{\"a}t Bonn,             
                   Bonn, Germany}                                             
\centerline{$^{22}$Physikalisches Institut, Universit{\"a}t Freiburg,         
                   Freiburg, Germany}                                         
\centerline{$^{23}$Institut f{\"u}r Physik, Universit{\"a}t Mainz,            
                   Mainz, Germany}                                            
\centerline{$^{24}$Ludwig-Maximilians-Universit{\"a}t M{\"u}nchen,            
                   M{\"u}nchen, Germany}                                      
\centerline{$^{25}$Fachbereich Physik, University of Wuppertal,               
                   Wuppertal, Germany}                                        
\centerline{$^{26}$Panjab University, Chandigarh, India}                      
\centerline{$^{27}$Delhi University, Delhi, India}                            
\centerline{$^{28}$Tata Institute of Fundamental Research, Mumbai, India}     
\centerline{$^{29}$University College Dublin, Dublin, Ireland}                
\centerline{$^{30}$Korea Detector Laboratory, Korea University,               
                   Seoul, Korea}                                              
\centerline{$^{31}$SungKyunKwan University, Suwon, Korea}                     
\centerline{$^{32}$CINVESTAV, Mexico City, Mexico}                            
\centerline{$^{33}$FOM-Institute NIKHEF and University of                     
                   Amsterdam/NIKHEF, Amsterdam, The Netherlands}              
\centerline{$^{34}$Radboud University Nijmegen/NIKHEF, Nijmegen, The          
                  Netherlands}                                                
\centerline{$^{35}$Joint Institute for Nuclear Research, Dubna, Russia}       
\centerline{$^{36}$Institute for Theoretical and Experimental Physics,        
                   Moscow, Russia}                                            
\centerline{$^{37}$Moscow State University, Moscow, Russia}                   
\centerline{$^{38}$Institute for High Energy Physics, Protvino, Russia}       
\centerline{$^{39}$Petersburg Nuclear Physics Institute,                      
                   St. Petersburg, Russia}                                    
\centerline{$^{40}$Lund University, Lund, Sweden, Royal Institute of          
                   Technology and Stockholm University, Stockholm,            
                   Sweden, and}                                               
\centerline{Uppsala University, Uppsala, Sweden}                              
\centerline{$^{41}$Physik Institut der Universit{\"a}t Z{\"u}rich,            
                   Z{\"u}rich, Switzerland}                                   
\centerline{$^{42}$Lancaster University, Lancaster, United Kingdom}           
\centerline{$^{43}$Imperial College, London, United Kingdom}                  
\centerline{$^{44}$University of Manchester, Manchester, United Kingdom}      
\centerline{$^{45}$University of Arizona, Tucson, Arizona 85721, USA}         
\centerline{$^{46}$Lawrence Berkeley National Laboratory and University of    
                   California, Berkeley, California 94720, USA}               
\centerline{$^{47}$California State University, Fresno, California 93740, USA}
\centerline{$^{48}$University of California, Riverside, California 92521, USA}
\centerline{$^{49}$Florida State University, Tallahassee, Florida 32306, USA} 
\centerline{$^{50}$Fermi National Accelerator Laboratory,                     
            Batavia, Illinois 60510, USA}                                     
\centerline{$^{51}$University of Illinois at Chicago,                         
            Chicago, Illinois 60607, USA}                                     
\centerline{$^{52}$Northern Illinois University, DeKalb, Illinois 60115, USA} 
\centerline{$^{53}$Northwestern University, Evanston, Illinois 60208, USA}    
\centerline{$^{54}$Indiana University, Bloomington, Indiana 47405, USA}       
\centerline{$^{55}$University of Notre Dame, Notre Dame, Indiana 46556, USA}  
\centerline{$^{56}$Purdue University Calumet, Hammond, Indiana 46323, USA}    
\centerline{$^{57}$Iowa State University, Ames, Iowa 50011, USA}              
\centerline{$^{58}$University of Kansas, Lawrence, Kansas 66045, USA}         
\centerline{$^{59}$Kansas State University, Manhattan, Kansas 66506, USA}     
\centerline{$^{60}$Louisiana Tech University, Ruston, Louisiana 71272, USA}   
\centerline{$^{61}$University of Maryland, College Park, Maryland 20742, USA} 
\centerline{$^{62}$Boston University, Boston, Massachusetts 02215, USA}       
\centerline{$^{63}$Northeastern University, Boston, Massachusetts 02115, USA} 
\centerline{$^{64}$University of Michigan, Ann Arbor, Michigan 48109, USA}    
\centerline{$^{65}$Michigan State University,                                 
            East Lansing, Michigan 48824, USA}                                
\centerline{$^{66}$University of Mississippi,                                 
            University, Mississippi 38677, USA}                               
\centerline{$^{67}$University of Nebraska, Lincoln, Nebraska 68588, USA}      
\centerline{$^{68}$Princeton University, Princeton, New Jersey 08544, USA}    
\centerline{$^{69}$State University of New York, Buffalo, New York 14260, USA}
\centerline{$^{70}$Columbia University, New York, New York 10027, USA}        
\centerline{$^{71}$University of Rochester, Rochester, New York 14627, USA}   
\centerline{$^{72}$State University of New York,                              
            Stony Brook, New York 11794, USA}                                 
\centerline{$^{73}$Brookhaven National Laboratory, Upton, New York 11973, USA}
\centerline{$^{74}$Langston University, Langston, Oklahoma 73050, USA}        
\centerline{$^{75}$University of Oklahoma, Norman, Oklahoma 73019, USA}       
\centerline{$^{76}$Oklahoma State University, Stillwater, Oklahoma 74078, USA}
\centerline{$^{77}$Brown University, Providence, Rhode Island 02912, USA}     
\centerline{$^{78}$University of Texas, Arlington, Texas 76019, USA}          
\centerline{$^{79}$Southern Methodist University, Dallas, Texas 75275, USA}   
\centerline{$^{80}$Rice University, Houston, Texas 77005, USA}                
\centerline{$^{81}$University of Virginia, Charlottesville,                   
            Virginia 22901, USA}                                              
\centerline{$^{82}$University of Washington, Seattle, Washington 98195, USA}  
}                                                                             
%end                                                                          

% \input list_of_visitor_addresses_r2.tex

\date{February 9, 2007}

\begin{abstract}
% remove the space for publication
\vspace*{1.0cm}
%%%%%%%%%%%%%%%%%%%%%%%%%%%%%%%%%%%%%%%%%%%%%%%%%%%%%%%%%%%%%%%%%%%%%%%%%%%%%%%%

A measurement of the top quark mass using events with one
charged lepton, missing transverse energy, and jets in the final
state, collected by the D0 detector from \ppbar collisions at
$\sqrt{s}$ = 1.96 TeV at the Fermilab Tevatron collider, is presented.
A constrained fit is used to fully reconstruct the kinematics of the events. For every 
event a top quark mass likelihood is calculated taking into account all
possible jet assignments and the probability that an event is signal
or background.  Lifetime-based
identification of $b$ jets is employed to enhance the separation between
\ttbar signal and background from other physics processes and
to improve the assignment of the observed jets to the quarks in the
\ttbar hypothesis.
We extract a multiplicative jet energy scale factor $JES$ in-situ, greatly reducing the systematic effect related to the jet energy measurement.    
In a data sample with an integrated luminosity of 425 \ipb, we observe 230 candidate events, with an estimated background of 123 events, and  measure  $ m_t = 173.7 \pm 4.4 \mbox{ (stat + JES)}^{+2.1}_{-2.0} \mbox{ (syst) GeV}$.
This result represents the first application of the Ideogram technique to the measurement of the top quark mass
in lepton+jets events.

\end{abstract}

% activate the following line for publication

\maketitle

%\newpage
%\cleardoublepage

\section{Introduction}
\label{sec:intro}

The top quark is by far the heaviest known fermion. Its discovery in
1995~\cite{discovery} confirmed the structure of the standard model, while its strikingly large mass compared to other fermions highlights remaining open questions
related to the large range of quark and lepton masses and the
precise mechanism of electroweak symmetry breaking that explains fermion masses in the theory.
Within the framework of the standard model, the top quark
mass is related to the Higgs boson mass and the $W$ boson mass through
radiative corrections. A precise measurement of the top quark mass helps
to constrain the standard model and to predict the mass of the Higgs
boson~\cite{EWWG}. At the same time it provides a challenge to 
the standard model with increased precision and distinguishes possible extensions of it.

%The Standard Model has been very succesful in describing accurately
%all known elementary particles and their interactions. However, it
%does not predict the values of the fermion masses nor does it offer an
%explanation of the wide variation in masses between the lightest
%fermions and the top quark. As the heaviest particle known to date,
%the top quark plays a special role due to its strong coupling to the
%Higgs field, and possibly a handle to understand more about the
%mechanism of electroweak symmetry breaking. But above all, a precise
%measurement of the top quark mass helps to constrain the Standard
%Model through loop corrections and predict the mass of the Higgs
%boson. It also allows to challenge Standard Model and distinguish
%possible extensions beyond it. 

At the Fermilab Tevatron Collider, which collides protons and antiprotons with a center-of-mass energy of
1.96~TeV, the top quark is predominantly produced in \ttbar
pairs through \qqbar annihilation ($\approx 85\%$) and gluon-gluon
fusion. In the framework of the standard model, the top quark decays 
almost exclusively to a $b$ quark and a $W$ boson. Thus
the final state topology of a \ttbar event is determined by the 
decay modes of the two $W$ bosons. The analysis presented in this
paper uses the lepton+jets (\ljets) channel, where one $W$ boson 
decays hadronically and the other $W$ boson decays leptonically to a
muon or an electron and the corresponding (anti)neutrino. Tau leptons are
not explicitly reconstructed in the analysis. Throughout this paper, 
charge conjugate modes are implicitly included.

The \ljets topology combines a sizable branching fraction with a
 striking signature of the isolated energetic lepton and large 
missing  transverse energy from the escaping neutrino.
The background from $W$+jets and QCD multijet
events is manageable. This means that the \ljets channel is
particularly suited for studies of top quark properties, and it has provided
the most precise measurements of the top quark mass to
 date~\cite{run1templmass,massCDF,ME,otherLJprecise, MEnew}. 

Nonetheless, serious challenges exist. A direct measurement of the top quark
mass requires that the kinematics of the event are fully reconstructed,
including the momentum of the neutrino. The signal events need to be
separated from backgrounds in a manner that does not bias the mass
measurement. Furthermore, with four jets in the final state, the
assignment of jets to the original top quark decay products gives a
twelve-fold ambiguity. Finally, a proper calibration of the jet energy
scale is crucial. In previous measurements, this was the 
dominant systematic uncertainty.

% new version after comment from Pushpa
Early measurements of the top quark mass~\cite{run1templmass} (and some recent analyses~\cite{massCDF}) used a constrained fit to reconstruct the kinematics of the event, choosing one jet assignment based on the quality of the fit.
A distribution of some event variable strongly correlated to the top mass, typically the fitted top mass, was plotted for data events.
This distribution was then compared to distributions based on Monte Carlo simulation generated for different top quark masses to determine the value of the top quark mass that best agrees with the data.
In the case of the D0 analysis~\cite{run1templmass}, a multi-variate discriminant that separates signal from background was also used in a two-dimensional likelihood fit to the Monte Carlo reference distributions.
However, in these analyses, only a certain amount of information per
event is used in the final fit.
%, which becomes an issue given the complexity of \ljets events.

%Earlier measurements of the top quark mass~\cite{run1templmass} (and
%some recent analyses~\cite{massCDF})
%used a kinematic fit followed by a Monte Carlo template fit.
%The limitation of this approach, however, is that only a certain amount of
%information per event is used in the final fit, which becomes an issue
%given the complexity of \ljets events. 

The D0 Matrix Element analysis~\cite{ME} demonstrated for the first time that
the statistical precision of the measurement can be greatly enhanced
by constructing event-by-event likelihoods that reflect the
full ambiguity of the events. A dramatic improvement was achieved,
albeit at the cost of computationally intensive methods.

The analysis presented in this paper uses the Ideogram technique. 
This method is based on a constrained kinematic fit and strives to obtain 
a similar improvement in statistical precision as the Matrix Element
analysis with minimal additional
computation. The constrained fit is used to determine the kinematics of the
events and to improve their reconstruction beyond the detector
resolution. A top quark mass likelihood is derived for every event including all
possible assignments of jets to quarks in the \ttbar hypothesis, and
taking into account the possibility that the event is background. The
top quark mass is extracted through a combined likelihood fit including all
events. This approach is very similar to the 
Ideogram technique used by the DELPHI experiment to measure the $W$
boson mass at the CERN LEP collider~\cite{DELPHI}. Also there the
different possible jet permutations lead to an ambiguity in the mass
fit which is reflected in the event likelihood as the sum of Gaussian
resolution functions. The similarity with the ideogram plots used by the
Particle Data Group~\cite{PDG} to visualize a set of measurements is
what gave the method its name.
This is the first time the method is used to determine the top quark mass
% leave it out for now:
in the \ljets channel. Recently, it has also been applied to the
all-hadronic decay channel~\cite{CDFIdeogram}.

The free parameters in the fit are the top quark mass, the \ttbar signal
fraction in the sample, and an overall jet energy scale (JES)
factor. Including the JES factor as a free parameter in the fit
greatly reduces the systematic uncertainty related to the jet energy 
scale calibration~\cite{massCDF,MEnew}. We employ the tagging of $b$
jets, i.e., $b$ tagging, to enhance the separation between
signal and background from other physics processes. The
 $b$ tags also help to better distinguish between correct and
 wrong jet assignments in the likelihood. Events with and without $b$
 tags are included in the overall likelihood fit. 

This paper is organized as follows:
Sections~\ref{sec:detector} and~\ref{sec:reco} describe the D0 Run II detector and the event reconstruction, respectively.
Sections~\ref{sec:sample} to~\ref{sec:fit} describe the data and simulation samples used and outline the event selection.
In Sec.~\ref{sec:disc}, the sample composition is estimated using topological and $b$ tagging information.
Section~\ref{sec:ID} describes in detail the calculation of the Ideogram likelihood and the Monte Carlo calibration procedure.
The method is applied to data in Sec.~\ref{sec:data} and the systematic uncertainties are discussed in Sec.~\ref{sec:systuncs}.
Section~\ref{section:Xcheck} presents a cross-check of the JES calibration, followed by the conclusion in Sec.~\ref{sec:conclusions}.

\newcommand{\psgnM}{\ensuremath{P_{\rm sgn}}\xspace}
\newcommand{\pbkgM}{\ensuremath{P_{\rm bg}}\xspace}
\newcommand{\psgnD}{\ensuremath{P_{\rm sgn}}\xspace}
\newcommand{\pbkgD}{\ensuremath{P_{\rm bg}}\xspace}
\newcommand{\mmm}{\ensuremath{x_{\rm fit}}\xspace}
\newcommand{\barq}{\ensuremath{\bar{q}}\xspace}
\newcommand{\barb}{\ensuremath{\bar{b}}\xspace}

\section{THE D0 DETECTOR}
\label{sec:detector}

The Run II of the Fermilab Tevatron collider started in 2001 after
substantial detector upgrades following the first Tevatron collider
run in 1992-1996. The D0 Run II detector~\cite{run2det} consists of 
a magnetic central tracking system, 
composed of a silicon micro-strip tracker (SMT) and a central fiber 
tracker (CFT), both located within a 2~T superconducting solenoidal 
magnet. The SMT has approximately $800,000$ individual strips, 
with typical pitch of $50-80$ $\mu$m, and a design optimized for 
tracking and vertexing capabilities at pseudorapidities of $|\eta|<2.5$. 
The system has a six-barrel longitudinal structure, each with a set 
of four layers arranged axially around the beam pipe, and interspersed 
with 16 radial disks. The CFT has eight thin coaxial barrels, each 
supporting two doublets of overlapping scintillating fibers of 0.835~mm 
diameter, one doublet being parallel to the beam axis, and the 
other alternating by $\pm 3^{\circ}$ relative to the axis. Light signals 
are transferred via clear fibers to solid-state photon counters (VLPCs) 
that have $\approx 80$\% quantum efficiency.

Central and forward preshower detectors located just outside of the superconducting coil (in front of the calorimetry) are constructed of several layers of extruded triangular scintillator strips that are read out using wavelength-shifting fibers and VLPCs.
The next layer of  detection involves three liquid-argon/uranium calorimeters: a central section (CC) covering approximately $|\eta| < 1.1$, and two end calorimeters (EC) that extend coverage to $|\eta|\approx 4.2$, all housed in separate cryostats~\cite{run1det}.
The calorimeter consists of an electromagnetic (EM) section followed by fine and coarse hadronic sections with modules assembled in a projective geometry to the interaction region.
In addition to the preshower detectors, scintillators between the CC and EC cryostats provide sampling of developing showers for $1.1<|\eta|<1.4$.

A muon system~\cite{run2muon} resides beyond the calorimetry and consists of a 
layer of tracking detectors and scintillation trigger counters 
before 1.8~T iron toroids, followed by two similar layers after
the toroids. Tracking for $|\eta|<1$ relies on 10~cm wide drift
tubes~\cite{run1det}, while 1~cm mini-drift tubes are used for
$1<|\eta|<2$.

%Luminosity is measured using plastic scintillator arrays located in front 
%of the EC cryostats, covering $2.7 < |\eta| < 4.4$. 
%A forward-proton 
%detector, situated in the Tevatron tunnel on either side of the 
%interaction region, consists of a total of 18 Roman pots used for 
%measuring high-momentum charged-particle trajectories close to the 
%incident beam directions.
Trigger and data acquisition systems are designed to accommodate  the high luminosities of Run II.
Based on preliminary information from tracking, calorimetry, and muon systems, the output of the first level of the trigger is used to limit the rate for accepted events to approximately $2$~kHz.
At the next trigger stage, with more refined information, the rate is reduced further to about 1~kHz.
These first two levels of triggering rely mainly on hardware and firmware.
The third and final level of the trigger, with access to all of the event information, uses software algorithms and a computing farm, and reduces the output rate to about 50~Hz, which is written to tape.

\section{EVENT RECONSTRUCTION}
\label{sec:reco}

This section summarizes the offline event reconstruction. We
use a right-handed Cartesian coordinate system with the $z$ axis defined by the
direction of the proton beam, the $y$ axis pointing vertically upwards
and the $x$ axis pointing out from the center of the accelerator
ring. The origin is at the center of the detector. 
The polar angle $\theta$ is defined with respect to the positive
$z$ axis and $\phi$ is the azimuthal angle from the $x$ axis in the transverse $xy$
plane. The pseudorapidity $\eta$ is defined as $\eta
\equiv -{\rm ln}({\rm tan}(\theta / 2))$.

\subsection{Tracks and event vertex}
Tracks are reconstructed from the hit information in the SMT and CFT.
A Kalman filter \cite{kalman} is used to fit track candidates found by
a road-based algorithm or a technique searching for clusters of track
parameters formed by tracker hits.
Using a vertex search procedure \cite{btagxsecprd}, a list of
reconstructed primary vertices is returned.
The primary event vertex for the \ttbar reconstruction 
is chosen from this list based on the $p_T$ spectrum of the tracks associated with a given vertex.
Only vertices with at least three tracks associated with them are considered.

\subsection{Electrons}
\label{sec:electrons}
We reconstruct electrons using information from the calorimeter and
the central tracker.
Clusters of EM calorimeter cells (EM clusters) are built with a simple
cone algorithm using seeds of $E_T>1.5$~GeV and radius 
$\Delta R \equiv \sqrt{(\Delta \eta)^2 + (\Delta \phi)^2} = 0.2$.
An ``extra-loose'' electron is defined as an EM cluster with 90\% of its energy from the EM part of the calorimeter and isolated from hadronic energy depositions.
Its longitudinal and transverse energy profiles have to be consistent with expectations from simulation.
In addition, the electrons used in the final event selection are
required to match a track reconstructed in the central tracker and to pass an electron likelihood cut.
The likelihood is built from seven variables containing tracking and calorimeter information and is optimized to discriminate between true electrons and background.

\subsection{Muons}
\label{sec:muons}
Muons are reconstructed from the information in the muon system and
the central tracker.
We require a muon candidate to have hits in the muon detectors both inside and outside the toroid.
The timing information of the scintillator hits has to be consistent with that of a particle produced in a $p \bar p$ collision, thus rejecting cosmic muons.
The muon candidate track is then extrapolated to the point of closest
approach to the beam line, and matched to a track from the central
tracking system using a global track fit.
Muons must not be surrounded by activity in the tracker or calorimeter
and are required to be separated from reconstructed jets by $\Delta R>0.5$.

\subsection{Jet reconstruction and energy scale}
\label{sec:jet}
Jets are reconstructed from the calorimeter information using a cone
algorithm \cite{blazey} with radius $\Delta R = 0.5$. Only calorimeter cells with signal larger than $4\sigma$ above the average noise and adjacent cells with signal at least $2 \sigma$ above the noise are used.
The jets are required to be confirmed by independent calorimeter
trigger information and must be separated from an extra-loose electron by $\Delta R>0.5$.
The reconstructed jet energies $E^{\rm jet}_{\rm reco}$ are corrected
for an energy offset $E_{\rm off}$, energy response $R_{\rm cal}$, and out-of-cone
showering $C_{\rm cone}$, according to:
\begin{equation}
E^{\rm jet}_{\rm corr} = \frac{E^{\rm jet}_{\rm reco} 
- E_{\rm off}}{R_{\rm cal} \cdot C_{\rm cone}} \,.
\end{equation}
The offset correction is determined from events taken with a zero bias
trigger during physics data taking and accounts for noise, multiple
interactions, and energy pile-up. 
The response correction is derived from a high statistics
jet+photon sample by looking at the $p_T$ imbalance in these  events.
The photon energy scale is assumed to be equivalent to the well-known
electron energy scale as calibrated from $Z\rightarrow ee$ events. 
The showering correction accounts for energy that particles inside the
jet cone deposit outside the cone during the hadronic showering
process. 
Transverse jet energy profiles are studied to determine this correction.
Jets containing a muon within the jet cone are further corrected for
the momentum carried by the muon and the associated neutrino.
Since the method to extract the top quark mass is calibrated with
respect to the Monte Carlo simulation, it is important to determine
the relative jet energy scale $\cal S$ between data and the Monte
Carlo simulation,
\begin{equation}
\label{eq:dat3}
{\cal S} = \left\langle \frac{p^{\rm jet}_T - p^\gamma_T}{p^\gamma_T} \right\rangle_{\rm data} 
	 - \left\langle \frac{p^{\rm jet}_T - p^\gamma_T}{p^\gamma_T} \right\rangle_{\rm MC}.
\end{equation}
$\cal S$ is parameterized as a function of photon $p_T$ for several
bins in ($p_T$,$\eta$) space and is found to be flat within its uncertainties.
No corrections from this source are therefore applied.
Effects of a potential $p_T$ dependence are taken into account as a systematic uncertainty.
For the overall jet energy scale, a uniform factor, $\jes$, is
introduced  as a free parameter in the analysis.
This factor is fitted in situ, simultaneously with the top quark mass
in data by using information from the invariant mass of the
hadronically decaying $W$~bosons. 
For every event, this mass is constrained in the kinematic fit to be equal to the known value of the $W$ boson mass~\cite{PDG}.
The $\chi^2$ of the kinematic fit reflects the compatibility of the reconstructed jet energies with this constraint.
The likelihood is sensitive to the $\jes$ parameter through the $\chi^2$. 
%Thus the $\chi^2$ of the kinematic fit provides sensitivity to
%the JES parameter in the likelihood. 
The overall fit will give the maximum likelihood for the 
value of $\jes$ which optimizes, on average, the
compatibility between the reconstructed and fitted jet energies.

%The overall jet energy scaling factor $\jes$ affects which gives
%the best overall compatibility with this hypothesis is the corresponds to the jet energy for
%which this compatibility 

Apart from the $W$ boson mass information, no constraint on the 
overall energy scale is used in the top quark mass fit.
The jet energy scale measured in situ is
consistent with the result obtained from photon+jet studies (Sec.~\ref{section:Xcheck}).

%calibration 
%is consistent with the Jet Energy Scale that is measured in situ. 
%is used to estimate the residual
%systematic uncertainty on the top mass due to $p_T$ and $\eta$
%dependent variations due to non-uniform JES effects that are not
%absorbed by fitting an overall JES factor $\jes$ (see section xx).

%To be consistent with the Matrix Element analysis, the measurement of
%the average difference ${\cal S} (p_T)$ is not used as a constraint in
%the fit. 

The analysis is calibrated such that in pseudo-experiments with Monte Carlo events the average fitted $\jes$ value is equal to one.
A fitted value $\jes <$ 1 means that the jet energies in the sample considered are underestimated with respect to the reference Monte Carlo scale described above ($\jes <$ 1 is equivalent to ${\cal S} < 0$ when fitting the data sample).

\subsection{Missing transverse energy}
We identify neutrinos indirectly from the energy imbalance in the event.
The imbalance is reconstructed from the vector sum of the transverse energies in the calorimeter cells and the reconstructed muons.
Energies from the cells in the coarse hadronic portion of the calorimeter are only added if associated with a reconstructed jet.
The missing transverse energy, \MET, is corrected for the energy scale calibration of jets and electrons.

\subsection{$b$-jet identification}
We identify $b$ jets using a lifetime tagging algorithm (secondary
vertex tagger, SVT) based on the explicit reconstruction of a secondary vertex from the decay of a $b$-flavored hadron~\cite{SVT}.
We call $dca$ the distance of closest approach between a track and the beam line, with $\sigma(dca)$ being the uncertainty on $dca$.
After the reconstruction of the primary event vertex, we consider tracks with $dca/\sigma(dca)>3.5$ for the reconstruction of additional (secondary) vertices.
For a reconstructed secondary vertex, the transverse decay length
$L_{xy}$ with respect to the primary event vertex is computed.
A jet is tagged as a $b$ jet if a secondary vertex is reconstructed within $\Delta R<0.5$ of the jet with $L_{xy}/\sigma(L_{xy})>7.0$, where $\sigma(L_{xy})$ is the uncertainty on $L_{xy}$.
The $b$-jet tagging rate $\epsilon_b$ is measured in data using information from an independent $b$-tagging analysis that looks for the presence of a muon in the jet cone.

Light quark jets can also be tagged when a fake secondary
vertex is reconstructed due to track mis-measurements
and random overlaps of tracks.
This light jet tagging rate  $\epsilon_l$ is estimated from the rate
of secondary vertices with $L_{xy}/\sigma(L_{xy})<-7.0$ in a data
sample with predominantly light quark jets. Negative values of
$L_{xy}$ occur if the secondary vertex is on the opposite side of the
event vertex with respect to the jet and are a sign of
mis-measurement and resolution effects.
Misreconstructed vertices with negative and positive values of
$L_{xy}$ are expected to occur at the same rate.
Corrections for the contamination with heavy flavor and the presence of long lived particles are applied as determined from Monte Carlo simulation.
The $b$-jet and light-jet tagging rates are measured in data and are parametrized as a function of jet transverse momentum and pseudorapidity~\cite{btagxsecprd}.

%The tagging rates and $b$ jet identification are used to improve the separation between signal and background events and to better distinguish between `correct' and `wrong' jet assignment hypotheses.
%Events with 0, 1, or $\geq 2$ $b$ tagged jets are all used in a combined fit in this analysis.

\section{DATA SAMPLES}
\label{sec:sample}
This paper describes the analysis of data collected between April 2002 and August 2004, corresponding to an integrated luminosity of approximately 425 pb$^{-1}$.
For this analysis, the data sample was selected by triggering on a lepton and at least one additional jet in the events.
The specific trigger requirements are described in more detail in Ref.~\cite{btagxsecprd}.

The event selection requires an isolated lepton of transverse momentum $p_T > 20$~GeV, with a pseudorapidity $|\eta| < 1.1$ for electrons and $|\eta| < 2$ for muons.
Missing transverse energy \MET $> 20$~GeV is required as well as four or more jets with $p_T > 20$~GeV and $|\eta|<2.5$.
A $\Delta \phi$ cut between \MET and lepton momentum is imposed to exclude events where the transverse energy imbalance is caused by a poor measurement of the lepton energy. 
The position of the event vertex along the beam direction has to be within 60 cm of the center of the detector.
We select 246 candidate events.

%A QCD multijet background sample is also extracted from data by reversing the lepton isolation requirements in the selection.

A QCD multijet background sample is also extracted from data by
reversing the final lepton quality requirement. Leaving all other
event
selection cuts unchanged, the candidate isolated muon must fail to
be isolated from activity in the tracker or calorimeter (Sec.~\ref{sec:muons})
in the muon+jets channel. Similarly, in the electron+jets channel the
candidate electron must not be matched to a track or fail to pass the
electron likelihood cut (Sec.~\ref{sec:electrons}).

\section{SIMULATION}
\label{sec:sim}
Monte Carlo event generators are used to create large samples of simulated signal and background events.
These samples are used for the calibration of the central mass value
and the estimate of the uncertainty.
We use \alpgen~1.3~\cite{ALPGEN} to generate signal and \wjets
background events.
The underlying event and hadronization is simulated using \pythia~6.2~\cite{PYTHIA}.
Signal \ttbar events are generated at nine mass points with masses ranging from 150~GeV to 200~GeV.
The factorization and renormalization scales are set to $Q=m_t$ for
the \ttbar simulation and $Q^2=M_W^2+\sum(p_T^{\rm jet})^2$ for \wjets.
All events are passed through a full \geant-based \cite{geant} D0 detector simulation and reconstructed with the same software as the collider data.
Events are accepted according to the probability that a simulated event would pass the trigger requirements.
This probability is typically between 0.9 and 1.0. 
The same object and event selections as for the data samples are applied.
The simulation chain is tuned to reproduce resolutions of
reconstructed objects seen in the collider data.
%For the $W$+jets background sample we build the relative reconstructed jet flavor composition are shown in table \ref{tab:wjets}.

\section{KINEMATIC FIT AND FINAL EVENT SELECTION}
\label{sec:fit}

The kinematics of the events, including the undetected neutrino from the $W$ boson decay, are
reconstructed using the same kinematic constrained fit that was developed for the
Run~I analysis~\cite{run1templmass}. The resolutions of
muons, electrons and jets were updated for Run~II~\cite{MEnew,Whel,Charge}. 

In events with more than four jets, only the four jets with highest $p_T$ are considered as
possible candidates to be a light quark or $b$ quark in the \ttbar
hypothesis used in the constrained fit. 

All twelve possible assignments of jets to quarks are considered.
As a starting point for the kinematic fit, the unmeasured component of the
neutrino momentum parallel to the beam, $p^{\nu}_z$, is chosen such that the two top 
quarks are assigned equal mass. This yields a quadratic equation for $p^{\nu}_z$.
We use both solutions as input to the fit yielding twenty-four fit results per event.
Depending on the event kinematics and resolution effects, the
discriminant of the quadratic equation may be negative, in which case
the discriminant is forced to be zero.
Thus one or two solutions are always obtained.
If only one solution is available, we include the same fit result twice in the likelihood.
% replaced by above after comment by Greg L. 12/28/06
% When the event kinematics yield two 
% solutions for the component of the neutrino momentum parallel to the
% beam, both solutions are used as a starting point for the fit, giving a
% maximum of 24 different fit solutions per event.

For the kinematic fit, we relate the reconstructed jet energy to the unfragmented parton energy.
To this end, a jet-parton energy mapping is applied, which is the same in data and MC simulation.
The corrections depend on the flavor ($b$ quark or light quark) of the parent quark and therefore depend on the jet-to-parton assignment used. 
To derive the mapping functions, we use MC events where the
jets are unambiguously matched to the partons of the \ttbar decay and compare the jet energy to the MC generated parton energy.
The jet-parton mapping functions contain the $\jes$ parameter as a uniform multiplicative factor.

The kinematic fit is performed by minimizing a  $\chi^2$ 
subject to the kinematic constraints: 
$m(t\rightarrow \ell\nu b) = m(\bar t\rightarrow q\barq\barb)$, $m(\ell\nu)=M_W$, and $m(q\barq) = M_W$.
We use $M_W=80.4$~GeV~\cite{PDG}.
The minimization algorithm uses the method of Lagrange multipliers; the nonlinear constraint equations are solved using an iterative technique.
From the fit for each jet/neutrino solution $i$, we extract the mass $m_i$, the
estimated uncertainty on the fitted mass 
$\sigma_i$, and the goodness of fit $\chi^2_i$. The fit is 
repeated for different values of $JES$.
The JES parameter is varied in steps of 3\% in an interval of $\pm$15\% around
unity.
Only jet combinations for which the fit converges at all values of $\jes$ are used. 
This requirement is needed to prevent
discontinuities as a function of $\jes$ in the event likelihood.
%As a final step in the event
%selection only event are kept for which at least one jet combination
%has a $\chi^2_i(\jes) < 10$, for $\jes = 1$. 
The fitted mass
$m_i(\jes)$, estimated uncertainty $\sigma_i(\jes)$, and goodness of fit
$\chi^2_i(\jes)$ all depend on the JES parameter. In the following
this dependence is not shown explicitly, to improve readability.

The final selection requirement is that at least one jet/neutrino 
solution yields $\chi^2 < 10$ for the kinematic fit with
$\jes=1$. This cut reduces the number of events from 120 to 116 in the
electron+jets channel and from 126 to 114 in the muon+jets
channel. Most of the events removed by this cut are background events
or badly reconstructed \ttbar events that do not satisfy the \ttbar fit
hypothesis and do not carry useful information about the top quark
mass. The algorithmic efficiency of the kinematic fit is
excellent, as listed in Table~\ref{tab:efficiencies}.

\begin{table}[h]
\begin{center}
\caption{\rm The numbers of events and efficiencies for the electron+jets
  ($e$) and muon+jets ($\mu$) channel having at least one jet
  combination for which the fit converges at $\jes = 1$, without and with the 
  requirement on the maximum value of the $\chi^2$. Each
  column shows the relative efficiency with respect to the previous column. }
\begin{tabular}{rcr@{}lr@{}lr@{}l}
\hline \hline
 \multicolumn{8}{l}{Convergence of the kinematic fit:} \\
       &  before \hspace{.5cm} 
       & \multicolumn{2}{c}{converges, \hspace{5mm}}  
       & \multicolumn{2}{c}{$\chi^2 < 10$  \hspace{5mm}} 
       & \multicolumn{2}{c}{converges,  \hspace{5mm}} \\
       & fit
       & \multicolumn{2}{c}{$\jes = 1$  \hspace{5mm} } 
       & \multicolumn{2}{c}{$\jes = 1$  \hspace{5mm} } 
       & \multicolumn{2}{c}{all $\jes^*$ \hspace{5mm}} \\
\hline
\ttbar  \hfill $e$ & 9452 & 100. & 0\%  &   97. & 7\% &   100. & 0\% \\
      $\mu$      & 9265 & 99. & 8\% &     94. & 4\% &   100. & 0\% \\
\wjets  \hfill $e$ & 5163 & 100. & 0\%  &   94. & 2\% &   100. & 0\% \\
      $\mu$      & 5820 & 99. & 7\% &     89. & 9\% &   100. & 0\% \\
data \hfill  $e$   &  120 &   100. & 0\%  &   97. & 0\%  &   100. & 0\% \\
      $\mu$      &  126 &  100. & 0\%  &   91. & 0\%  &   100. & 0\% \\
\multicolumn{8}{l}{$^*$ for all values of $\jes$ in the fit range $0.85 < \jes < 1.15$}\\
\hline \hline
\end{tabular}
\label{tab:efficiencies}

\end{center}
\end{table}

\section{SAMPLE COMPOSITION}
\label{sec:disc}

In order to obtain a good separation between \ttbar signal and background events (mainly $W$+jets), a likelihood discriminant based on the `low-bias'
topological discriminant $D_{\rm LB}$, developed in
Run~I~\cite{run1templmass}, is used.
The $D_{\rm LB}$ discriminant was designed to have minimal correlation
with the top mass and is based on the following four
topological variables: \met, aplanarity, $H'_{T2}$, and  $K'_{T \rm min}$.
Aplanarity is defined as the smallest eigenvalue of the normalized laboratory-frame momentum tensor of the jets and the $W$ boson. 
$H'_{T2} \equiv \frac{H_{T2}}{H_{\parallel}}$ measures the event
centrality, where $H_{T2}$ is the scalar sum of the transverse momenta 
of the jets excluding the leading jet and $H_{\parallel}$ is the sum of
the magnitudes of the momentum components parallel to the beamline of
the jets, isolated lepton and neutrino. In this case the neutrino momentum
parallel to the beam is estimated requiring that the mass computed
from the measured lepton momentum, \met and unknown neutrino momentum
parallel to the beam is equal to the $W$ boson mass. If more than one
solution is found, the one smallest in absolute value is used.
The variable $K'_{T \rm min} \equiv \frac{\Delta R^{\rm min}_{ij} \cdot E_{T}^{{\rm lesser}\
    j}}{E_{T}^{W}}$ is a measure of the jet separation normalized by the transverse energy of the reconstructed $W$ boson.
$\Delta R^{\rm min}_{ij}$ is the smallest distance in $\eta-\phi$ space between any
    two of the four leading jets.
$E_{T}^{{\rm lesser}\ j}$ is the smaller of the two jet $E_T$s.
The transverse energy of the $W$ boson is defined as
$E^{W}_{T} \equiv \left|p_{T}^{\rm lepton}\right| + \left| \met \right|$.
These four variables are combined in a single discriminant variable
$D_{\rm LB}$ using the likelihood ratio procedure described in Ref.~\cite{run1templmass}.

For the analysis presented here, the low-bias discriminant $D_{\rm LB}$
($\equiv x_1$) was combined with a new variable called ``$p_T$-fraction''
and the number of $b$ tags to build a combined discriminant $D$. 
The $p_T$-fraction, defined as $ x_2 = (\sum_{\rm tracks\ in\ jets} p_T)/(\sum_{\rm all\ tracks} p_T)$, is the
$p_T$-weighted fraction of all tracks in the event that point to an
energy deposit defining a jet (with jet $p_T>$20~GeV with $|\eta| < 2.5$).
Only those tracks were considered that have a distance of closest
approach of less than 1 cm along the beam direction with respect to at least one of the primary vertices in the event.
In order to be included in the $p_T$ sum over tracks in a jet, a track was required to be within $\Delta R < 0.5$ from the jet axis.
This variable distinguishes clean events with nicely collimated jets from
events with broader jets and significant underlying hadronic activity.
Finally, $x_3$ is the number of $b$ tags. 
For each variable $x_i$, we use Monte Carlo simulation to determine
 the probability density functions
$s_i$, for \ttbar signal, and $b_i$, for \wjets background.
To a good approximation, these three
variables $x_i$ are uncorrelated, and the combined likelihood discriminant is 
derived as 
\begin{equation}
\label{eq:dat2}
D = \frac{\Pi_i s_i(x_i)/b_i(x_i)}{\Pi_i
  s_i(x_i)/b_i(x_i) + 1},
\end{equation}
thus combining event topology with a tracking-based jet shape and $b$ tag
information. This combined likelihood discriminant offers a much better
discrimination between \ttbar and background than does the low-bias
topological variable $D_{\rm LB}$ by itself, while maintaining its low
level of correlation with the fitted top quark mass (and therefore with the
jet energy scale).

\begin{figure*}
 \includegraphics[width=\linewidth]{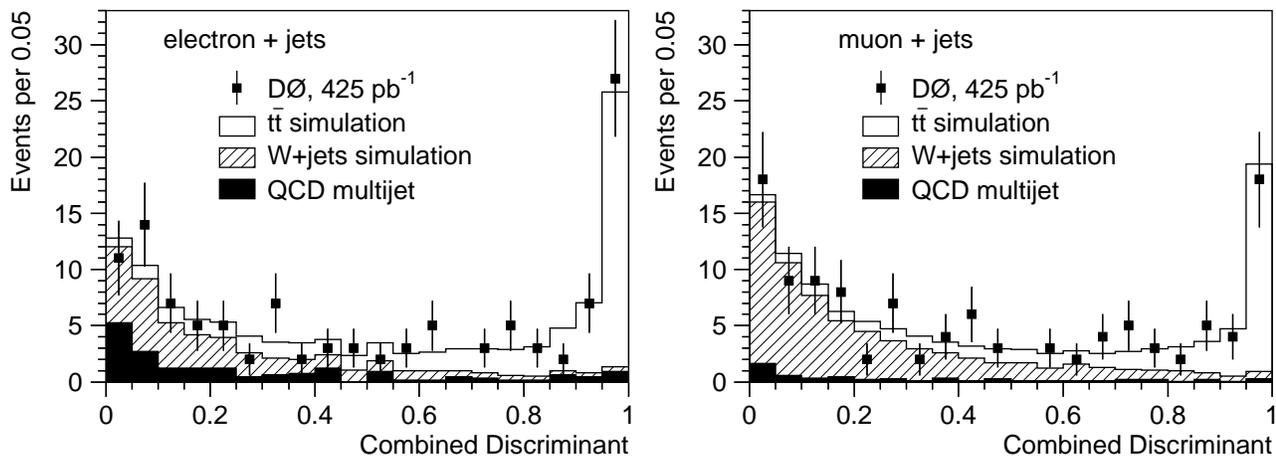}
 \caption{\label{fig:PurClosure} {Combined likelihood discriminant $D$
 in data and MC simulation in the electron+jets channel (left) and muon+jets channel
 (right). The \ttbar, \wjets and multijet contributions are normalized
 according to the fitted fractions.}}
\end{figure*}

Figure~\ref{fig:PurClosure} shows the distribution of the combined
discriminant $D$ obtained in the electron+jets and muon+jets channels.
The distribution observed in data is compared to a model consisting of simulated \ttbar
and \wjets events and the QCD multijets sample obtained
from data (Sec.~\ref{sec:sample}). 
%obtained from data by inverting the
%lepton isolation cut in the event selection. 
A likelihood fit is performed to determine the estimated fraction of \ttbar
events.
The fit results are shown in Table~\ref{tab:selection}.
In the fit, the ratio between the number of QCD and \wjets events was kept 
fixed at a value based on the estimate used in Refs.~\cite{topological,MEnew}.

%likelihood fit to \ttbar and W+jets.
%The closure plots all look good, showing no sign of
%correlations spoiling that assumption.

\begin{table}[h]
\begin{center}
\caption{ Composition of the 425 pb$^{-1}$
            data sample as determined by the likelihood fit. }
\begin{tabular}{lr@{ }lr@{ }l}
\hline \hline
%%% \multicolumn{3}{l}{estimated composition:} \\
  & \multicolumn{2}{c}{electron+jets} & \multicolumn{2}{c}{muon+jets}  
%no. of events observed in data & 116 & 114 \\
%\hline
\\ \hline
%\multicolumn{3}{l}{estimated sample composition:} \\
\ttbar     & \hspace*{3mm}61.5 & $ \pm$ 7.9 &  \hspace*{1mm} 45.6 & $\pm$  7.5 \vspace{.15cm}\\
\wjets   & 35.6 & $\pm$ 5.2 & 63.0 & $\pm$  6.9 \vspace{.15cm}\\
QCD multijet       & 18.9 & $\pm$ 2.7  & 5.4 & $\pm$  0.6   \vspace{.15cm}\\
% \hline
Total observed & \multicolumn{2}{c}{116} & \multicolumn{2}{c}{114} \\ 
\hline \hline
\end{tabular}
\label{tab:selection}

\end{center}
\end{table}

\section{The Ideogram Method}

\label{sec:ID}
%
%In this section the measurement of the top quark mass with the ideogram method is described.
%The approach is very similar to a technique, which was used by the
%DELPHI experiment~\cite{DELPHI} to extract the mass of the $W$~boson at
%LEP. As in the Matrix Element analysis~\cite{ME} a likelihood is calculated for
%%%each event as a function of the assumed top quark mass and jet energy
%scale taking into account all possible jet assignments and the
%probability that the event was signal or background. As in the Matrix
%Element method
%the likelihood is described as a convolution of a
%physics function and the detector resolution. The difference, however,
%is that in the Ideogram method a kinematic constrained fit is used to 
%describe the detector resolution, and the physics function is
%simplified to a relativistic Breit-Wigner describing the average of
%the invariant masses of the supposed top and anti-top quark that were
%produced in the event.

To maximize the statistical information on the top quark mass
extracted from the event sample, a likelihood to observe the
event is calculated for each event as a function of the assumed top 
quark mass \mtop, the jet energy scale parameter $\jes$, and the fraction of \ttbar events in the event sample, \ftop.
The likelihood is composed of two terms, describing the
hypotheses that the event is \ttbar signal or background:
%
%The probabilities from all events in the sample are then combined to
%obtain the sample probability as a function of assumed mass and jet
%energy scale, and the top quark mass measurement is extracted from
%this sample probability.
%
%two processes, \ttbar production and \wjets events, and can be written
%as:
\begin{multline}
  \label{eq:idevlik}
   {\cal L}_{\rm evt}\left(x;\mtop,\jes,\ftop\right) = 
                     \ftop \cdot \psgn\left(x;\,\mtop,\jes\right) 
   \\ + \left(1-\ftop\right) \cdot \pbkg\left(x;\,\jes\right) \,. 
\end{multline}
Here, $x$ denotes the full set of observables that characterizes the event, \ftop is the
signal fraction of the event sample, and \psgn and \pbkg are the
probabilities for \ttbar and \wjets production, respectively. 
%
%\begin{equation}
%  \label{eq:idevlik}
%	{\cal L}_{\rm evt}\left(x; m_{t},\jes\right) = 
%  = 
%  N_s \cdot {\cal L}_{\rm sig} \left(x;\,\mtop,\jes\right) 
%    +  N_b \cdot {\cal L}_{\rm bckg}\left(x;\,\jes\right)
%  \ .
%\end{equation}
%
%Here, $x$ denotes the kinematic variables of the event, $N_s$ and $N_b$ are the 
%number of signal and background events in the sample, and \psgn and \pbkg are the 
%probabilities for \ttbar and \wjets, respectively.
%
The contribution from QCD multijet events is comparatively small and expected to have a fitted mass shape very similar to that of \wjets events.
Therefore no explicit QCD multijet term is included in the likelihood.
The event observables $x$ can be divided into two groups.
One set is chosen to provide good separation between signal and background events while
minimizing the correlation with the mass information in the event.
These variables (topological variables and $b$ tagging) are used to construct a low-bias combined
discriminant $D$, as described in Sec.~\ref{sec:disc}.
The other event information used is the mass information \mmm from the constrained
kinematic fit, which provides the sensitivity to the top quark mass and jet
energy scale.
To good approximation $D$ is uncorrelated with \mmm and
with the jet energy scale. Thus the probabilities \psgn and
\pbkg can be written as the product of a probability to
observe a value $D$ and a probability to observe \mmm, as
%\begin{equation}
%  \label{eq:idfactorised}
%    P_{evt}\left(D,\m;\,\mtop,\jes,\ftop\right)
%  =
%                     \ftop \cdot \psgn\left(D\right)\psgn\left(\m;\,\mtop,\jes\right)
%    + \left(1-\ftop\right) \cdot \pbkg\left(D\right)\pbkg\left(\m;\,\jes\right)
%  \ .
%\end{equation}
\begin{equation}
  \label{eq:idsigfactorised}
    \psgn\left(x;\,\mtop,\jes\right) \equiv \psgn\left(D\right) \cdot \psgn\left(\mmm;\,\mtop,\jes\right)
\end{equation}
and
\begin{equation}
  \label{eq:idbgfactorised}
    \pbkg\left(x;\,\jes\right) \equiv \pbkg\left(D\right) \cdot \pbkg\left(\mmm;\,\jes\right)
\end{equation}
where $D$ is calculated for a JES parameter equal to 1. The normalized
probability  distributions of the discriminant $D$ for signal
$\psgn\left(D\right)$ and
background $\pbkg\left(D\right)$ are assumed to be independent of 
$JES$ and are obtained from Monte Carlo
simulation as discussed in Sec.~\ref{sec:disc}. They correspond to
parameterized versions of the Monte Carlo templates shown in
Fig.~\ref{fig:PurClosure}. The reconstruction
of the signal and background probabilities for the mass information
\mmm is explained in Sec.~\ref{sec:psig}. The mass information in the event \mmm
consists of all fitted masses $m_i(\jes)$, estimated uncertainties $\sigma_i(\jes)$, and goodnesses-of-fit $\chi^2_i(\jes)$ obtained from the
kinematic fit.

\subsection{Calculation of signal and background probability}
\label{sec:psig}
The signal and background probabilities are calculated as a sum over all twenty-four possible jet/neutrino solutions.
Without $b$ tagging, the relative probability for each of the solutions $i$ to be correct depends only on the $\chi^2_i$ for the corresponding fit and is proportional to ${\rm exp}(-\frac{1}{2} \chi^2_i)$.
To further improve the separation between correct and incorrect jet
assignments, $b$ tagging is used. 
If one or more jets in the event are $b$ tagged, an additional relative
weight $w_{{\rm btag},i}$ is assigned, representing the probability that
the observed $b$ tags are compatible with the jet assignment assumed for that
particular jet permutation:
\begin{equation}
  \label{eq:id1}
 w_{{\rm btag},i} = \prod_{j=1,n_{\rm jet}} p_i^j ,
\end{equation}
where $p_i^j$ can either be $\varepsilon_l$, (1-$\varepsilon_l$), $\varepsilon_b$,
or (1-$\varepsilon_b$), depending on the assumed flavor of the jet
(light or $b$) and whether or not that particular jet is tagged. 
For this purpose the jets from the hadronic W boson decay are
always assumed to be light quark ($u$, $d$, $s$) jets. In the calibration of
the analysis (see Sec.~\ref{section:MC}), however, the fraction of $W \rightarrow
c\bar{s}$ decays and the higher tagging rate for $c$ quark jets are
taken into account. The tagging rates for light and $b$ quark jets $\varepsilon_l$ and
$\varepsilon_b$ are used as parameterized functions of jet
$p_T$ and $\eta$.
The jet $p_T$ is based on the reconstructed jet energy for $\jes = 1$, consistent with the jet energy scale for which the tagging rate functions are derived from data~\cite{SVT}. 
Thus, the weight assigned to each jet combination becomes 
\begin{equation}
  \label{eq:id2}
w_i = {\rm exp}(-\frac{1}{2} \chi^2_i)  \cdot w_{{\rm btag},i} .
\end{equation}
The mass-dependent signal probability in Eq.~\ref{eq:idsigfactorised} is calculated as 
\begin{multline}
  \label{eq:id3}
\psgn\left(\mmm;\,\mtop,\jes\right) \equiv \\ \sum^{24}_{i=1} w_i \Biggl[
f^{\rm ntag}_{\rm correct} \cdot \int^{m_{\rm min}}_{m_{\rm max}}
 {\bf G}(m_i,m',\sigma_i) \cdot {\bf BW}(m',m_{t}) dm' 
   \\ + (1-f^{\rm ntag}_{\rm correct})\cdot {S^{\rm
    ntag}_{wrong}}(m_i,m_{t}) \Biggr] .
\end{multline}
The signal term consists of two parts: one part describes the
compatibility of the solution with a certain value of
the top quark mass, assuming that it is the correct solution. It takes
into account the estimated mass resolution $\sigma_i$ for each jet
permutation.
The second part of the signal term describes the expected
shape of the mass spectrum for the ``wrong'' jet assignments, which also
depends on the top quark mass.
The ``correct'' solution part is given by a convolution of a Gaussian
resolution function ${\bf G}(m_i,m',\sigma_i)$  and a relativistic
Breit-Wigner ${\bf BW}(m',m_{t})$. The
Gaussian function describes the experimental resolution. 
The relativistic Breit-Wigner represents the expected distribution of
the average invariant mass $m'$ of the top and anti-top quark
in the event for a given top quark mass $m_{t}$. The width of the
Breit-Wigner is set to the standard model value of the top decay
width~\cite{PDG}. 
The ``wrong'' permutation signal
shape ${S^{\rm ntag}_{wrong}}(m_i,m_{t})$ is obtained from MC simulation
using a procedure described in Sec.~\ref{section:combi}.
These two terms are assigned relative weights depending on 
$f^{\rm ntag}_{\rm correct}$, which represents the relative
probability that the weight is assigned to the correct jet
permutation. For well-reconstructed events with exactly 4 jets, 
this probability is approximately 39\% if $b$ tagging is
not used.
For 4-jet events with 0, 1, or $\geq 2$ tagged jets, the values $f^0_{\rm correct}$=0.45, $f^1_{\rm correct}$=0.55, and
$f^2_{\rm correct}$=0.65 are used. 
For 5-jet events smaller fractions are used: 0.15, 0.30, and 0.40 for events
with 0, 1, or $\geq 2$ $b$ tagged jets respectively. Ensemble tests
(see Sec.~\ref{section:MC}) confirm that these values result in a pull width for the 
mass close to unity for the different tagging multiplicities. 

\begin{figure*}
\centerline{
  \includegraphics[width=\linewidth]{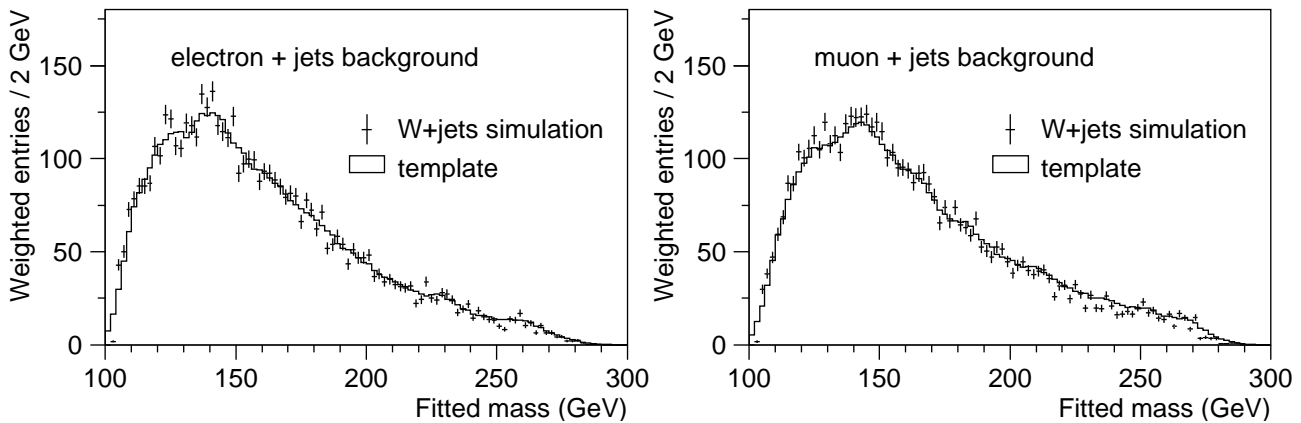}}
\caption{
\label{fig:bgshape} 
Histograms showing the background shape from a weighted sum (see text) of all twenty-four masses from each
event from the {\wjets} background sample (points with error bars), for the electron+jets
channel (left) and muon+jets (right).
The histograms show the shapes that are used in the likelihood.
To reduce statistical fluctuations, the shapes are calculated as the average value 
in a sliding window of $\pm$5~GeV around each fitted mass.}
\end{figure*}

The background term in Eq.~\ref{eq:idbgfactorised} is calculated as:
\begin{equation}
  \label{eq:id4}
\pbkg\left(\mmm;\,\jes\right) \equiv \sum^{24}_{i=1} w_i \cdot BG(m_i) ,
\end{equation}
where the background shape $BG(m)$ is the shape of the fitted mass spectrum for \wjets events.
To obtain $BG(m)$, the kinematic fit (with $\jes$ equal to unity) is applied to simulated
\wjets events and the fitted masses $m_i$ for all possible jet/neutrino solutions $i$ are plotted.
 All entries are weighted according to the permutation weight $w_i$ defined in
Eq.~\ref{eq:id2}.
The shapes of $BG(m)$ used in the analysis are shown in Fig.~\ref{fig:bgshape}. 

The Breit-Wigner and ``wrong'' permutation signal shape are normalized to unity within
the integration interval of $m_{\rm min}=100$~GeV to $m_{\rm max}=300$~GeV.
This interval is chosen to be large enough so as not to bias the mass in the region of
interest. 
%
%The normalization of the background
%shape $BG(m)$ is calibrated to
%minimize the offset in the fitted purity, in the region of interest around the
%expected signal fraction of 0.45.

The normalization of the background shape $BG(m)$ 
is chosen such that the fitted signal fraction \ftop reproduces
the true \ttbar fraction in ensemble tests (see Sec.~\ref{section:MC}) containing \ttbar and
\wjets events. The mass fit tends to underestimate \ftop, due to 
the presence of \ttbar events that are misreconstructed or affected by
energetic gluon radiation and resemble \wjets events in the fact that
their topology does not conform to the \ttbar hypothesis in the kinematic fit. 
A constant normalization factor of 1.15 is found to reduce the offset
in \ftop to less than 1\% both in the electron+jets and the muon+jets channel.
%
%%%%The event purity is $P_{\rm evt} = \frac {\ftop \cdot \frac{P_s}{P_b}}{1 + \ftop \cdot \frac{P_s}{P_b}}$.
%
The jet energy scale parameter is varied before performing the constrained fit by 
scaling all jet energies by a constant factor.
The event likelihoods are recalculated for each different value of the JES parameter. 
Since the constrained fit uses a $W$~boson mass constraint,
the $\chi^2$ in the fit will be best when the invariant mass of the
jets from the hadronically decaying $W$~boson is closest (on average) to the
known $W$~boson mass.
Additional sensitivity to the jet energy scale comes
from the shape of the fitted mass distribution in background events. For the proper
jet energy scale the spectrum will agree best with the background
shape included in the background term in the likelihood.

\subsection{Determination of the wrong-permutation signal shape}
\label{section:combi}

The convolution of Gaussian detector resolution and a Breit-Wigner,
used in the signal term of the likelihood, implicitly assumes that the
correct jet assignment is chosen.
To describe the contribution from wrong jet assignments, a separate term is added to the signal part of the likelihood.
%Since the 'wrong' combinations of jet assignments to the \ttbar decay partons do 
%not contain the correct information of the measured quantities, these
%combinations are treated separately in the likelihood.
To obtain the fitted mass spectrum of the wrong permutation signal, samples of parton-matched
\ttbar events are used in which all quarks are matched to jets.
The fitted mass spectrum is plotted including all jet permutations {\em except}
the correct solution (excluding both neutrino solutions corresponding
to the correct jet permutation). Each entry is weighted according to
the permutation weight assigned in the Ideogram likelihood.
Samples of different generated top quark masses are used.
%For each mass, the weighted sum of the correct and wrong solutions was normalized to a fixed number,
%while the ratio of correct and wrong solutions was maintained.
For each mass, the weighted sum of wrong solutions is fitted with a double
Gaussian.
%As shown in Figure~\ref{fig:combi1} this gives a reasonable description at all masses.
%As shown in Figure~\ref{fig:combi2} the correct solutions can be fitted with single Gaussians.
The fitted parameters for correct solutions and for the wrong
permutation signal show a linear behavior as a function of the
top quark mass. The fitted parameters are given in
Table~\ref{tab:combishapes}. Since the permutation weights change 
when $b$ tagging is included, this exercise is repeated for events 
with 0 tags, 1 tag, and 2 or more tags.

%, as shown in Figure~\ref{fig:combi3}, which also shows the linear dependences
%fitted.
%This procedure was repeated for each value of the JES scaling factor
%used (ranging from 0.85 to 1.15 in steps of 0.03). 
%As shown in Figure~\ref{fig:combijes} the bias and slope of the linear fits mentioned before show a nice linear dependence as function of the JES factor.

\begin{figure}
\ifthenelse{\lengthtest{\linewidth < 10cm}}
{ \includegraphics[width=\linewidth]{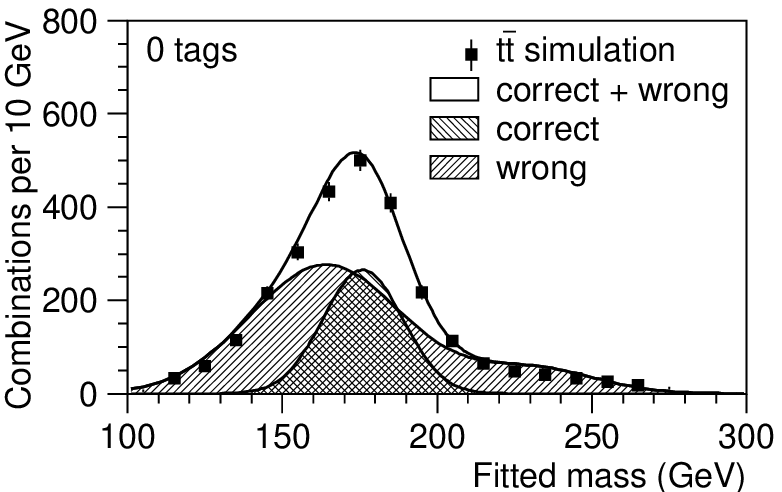}\\
  \includegraphics[width=\linewidth]{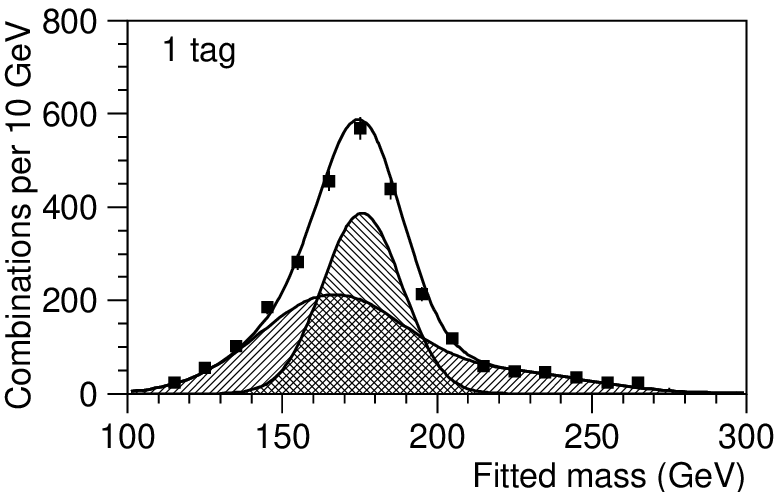}\\
  \includegraphics[width=\linewidth]{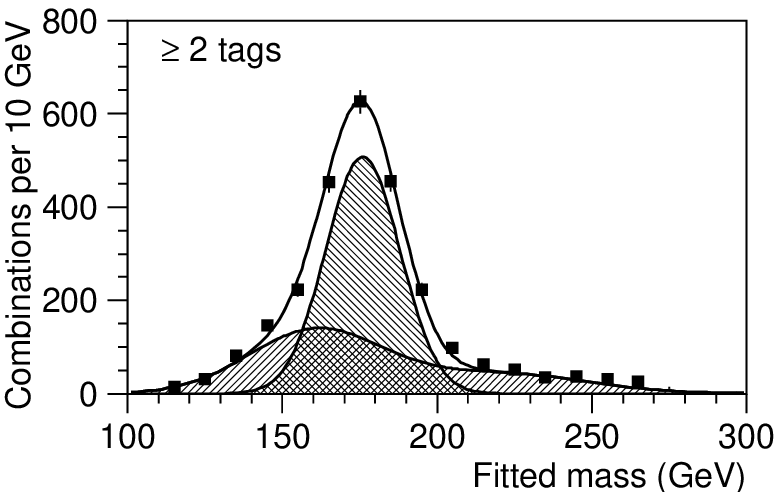}
}
{  \includegraphics[width=10cm]{figuresid/all_shapes_0tag.eps}\\
  \includegraphics[width=10cm]{figuresid/all_shapes_1tag.eps}\\
  \includegraphics[width=10cm]{figuresid/all_shapes_2tag.eps}
}
\caption{\label{fig:combi_vs_ntag} {Prediction of the shapes of the fitted
    mass distribution for the wrong and the correct permutations 
    (hatched) and the sum of the two
    (black line) using the fitted parameters
    shown in Table~\ref{tab:combishapes}. The sum of the two is
    compared to 
    the simulated data
    containing a weighted sum of all solutions (correct and wrong),
    for the default jet energy scale and a generated top quark mass of
    175~GeV for events with 0 (upper), 1 (middle), or more than 1 (lower) $b$ tags. }}
\end{figure}

\begin{figure*}
  \includegraphics[width=\linewidth]{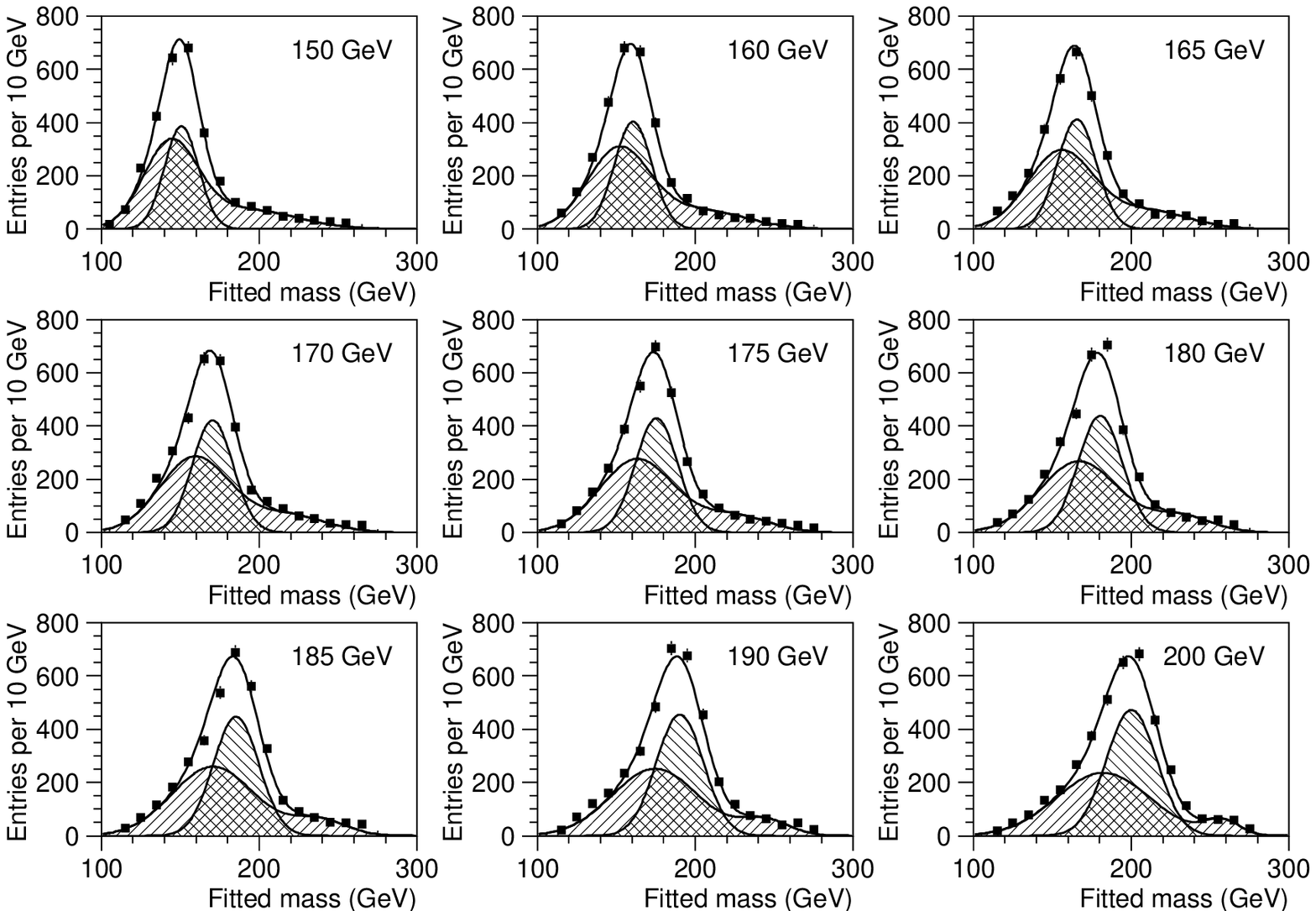} %\hglue -.5cm
\caption{\label{fig:combi_vs_mass} {Same as
    Fig.~\ref{fig:combi_vs_ntag}, for different values of the
    generated top quark mass, combining all events irrespective of the
    number of $b$ tags.}}
\end{figure*}

The linear fits are used to construct a 2-dimensional
wrong-permutation signal shape as a function of the fit mass and generated top
quark mass
${S^{\rm ntag}_{wrong}}(m_i,m_{t})$.
For each value of the
generated top quark mass, the shape as a function of fitted mass is described as the
sum of two Gaussians.
The resulting parameterizations are displayed as the
wrong-permutation shapes in Fig.~\ref{fig:combi_vs_ntag} and~\ref{fig:combi_vs_mass}.
%\begin{figure}[htb]
%  \includegraphics[width=16.cm]{figuresid/CombiShape.eps} 
%\caption{\label{fig:combi1} {The weighted sum of the fitted masses
%of all 'wrong' jet permutations, for MC samples with different
%generated top masses varying from 150 GeV (top left) to 200 GeV
%(bottom right) with default Jet Energy Scale (= 100\%). Events both from the e+jets and mu+jets channel are
%included. At each top mass the shape is fitted with a double Gaussian.}}
%\end{figure}
Also shown are the shapes of the correct jet assignments, determined in 
a similar fashion from parton-matched events using a single Gaussian. A linear
dependence of the parameters is found as a function of generated top quark
mass. The sum of the correct solutions and wrong solutions 
is compared to a weighted histogram of all fitted masses in \ttbar simulation. 
%The wrong
%solution shape is described by the double Gaussian with parameters
%shown in Table~\ref{tab:combishapes} while the correct solution shape
%was parametrized separately as a single Gaussian with parameters
%changing linearly as a function of the generated top quark mass. 
The parametrized functions give an adequate description of the overall
(wrong + correct) signal shape.
In Fig.~\ref{fig:combi_vs_ntag}, the corresponding distributions are shown for events
  with 0, 1, or 2 tags.  It is clearly visible how the fraction of the
  weight given to the correct solution improves when including $b$ tag
  information in the permutation weights.
In Fig.~\ref{fig:combi_vs_mass}, the nine distributions are shown for generated top
  quark masses ranging from 150~GeV to 200~GeV. 

\begin{table*}[htb]
\begin{center}
\caption{Parameters used to describe the background shapes
  (arbitrary normalization). For
  each case, the shape is described by the sum of two Gaussians $G(m_{\rm fit}$) $=
  a \cdot
  {\rm exp}\left[-({\rm \mu}-m_{\rm fit})^2/2\sigma^2\right]$, where the three
  parameters $a$, $\mu$, and $\sigma$ evolve linearly as a function of the
  generated top quark mass $m_t$ as $p0 + p1 \cdot (m_t - 175$ GeV$)$.}
\begin{tabular}{crrrrrrrrrrrrrrrrr}
\hline \hline
 & \multicolumn{6}{c}{ 0 tags } &  \multicolumn{6}{c}{1 tag}     & \multicolumn{5}{c}{$\geq$2 tags}\\
 & \multicolumn{3}{c}{Gaussian 1} &  \multicolumn{3}{c}{Gaussian 2} 
 & \multicolumn{3}{c}{Gaussian 1} &  \multicolumn{3}{c}{Gaussian 2} 
 & \multicolumn{3}{c}{Gaussian 1} &  \multicolumn{2}{c}{Gaussian 2}\\
Parameter & $p0$ & $p1$  && $p0$ & $p1$  && $p0$ & $p1$  && $p0$ & $p1$  && $p0$ & $p1$ && $p0$ & $p1$ \\
 \hline
$a$       &284.9&$-1.722$ && 51.72 &$-0.4199$ &\hglue 1.5cm &267.5 &$-1.0700$ &&68.08 & $-0.7129$ &\hglue 1.5cm & 235.5 & $-0.1662$ && 75.86 &$ -0.0415$ \\
$\mu$ & 161.7   &0.7383 &&223.1  & 1.242  && 162.6& 0.7859 &&220.1 & 1.400 && 166.2 & 0.6416 && 229.4 & 0.7454 \\
$\sigma$ &23.55 & 0.2392&& 22.94 &$ -0.2528$ &&23.27 & 0.2737 && 23.97 & $-0.4551$ && 25.80 & 0.1165 && 21.78 &$-0.2828$ \\
\hline \hline
\end{tabular} \\
\label{tab:combishapes}
\end{center}
\end{table*}

\subsection{Determination of JES offset correction}
\label{section:JESbias}

%The Ideogram likelihood described in the previous sections counts 
%on the kinematic fit to return the lowest $\chi^2$ (= best fit) when
%the invariant mass of the jets from the supposed $W$ boson in the 
%%event, scaled according to the overall 
%$\jes$ parameter, agrees best with the $W$ 
%boson mass used (=80.4 \GeV) in the fit (averaged over all
%events). This is expected to work fine for well-reconstructed \ttbar 
%events when the correct jet assigment is used. For wrong jet
%assignments, misreconstructed \ttbar events, or backgrounds from other
%physics processes there is no reason to assume that the fit procedure
%will yield an unbiased measurement of the $\jes$ parameter. Therefore, 

The likelihood fit relies on the invariant mass of the hadronically
decaying $W$~boson in the \ttbar events to set the jet energy
scale. It is designed to give an unbiased fit of the $\jes$
parameter in well-reconstructed \ttbar events when the correct jet
assignment is used. However, in a significant fraction of the events, the jets that are presumed to originate from the $W$~boson may not really come from a $W$~boson.
 Such cases include events other than \ttbar, as well as \ttbar events 
that are mis-reconstructed. In the presence of such events we can
expect an offset in the fitted JES parameter. 
The slope of the JES calibration curve (fitted JES parameter
as a function of the ``true'' JES) may also differ from unity. 
%Such effects are carefully investigated and corrected for.

Using the MC calibration procedure described in Sec.~\ref{section:MC}, we find that the presence of wrong jet
assignments and background events causes an offset of several percent
in the fitted JES parameter. A breakdown
of the different contributions to the JES offset and slope is shown in
Table~\ref{tab:JESslope}.
\begin{table}[h]
\begin{center}
\caption{The JES calibration slope and offset for different
  event samples are shown. The offset increases and the calibration slope
  becomes smaller when mis-reconstructed signal events or
  background events are added. The offset correction at the likelihood level (see
  text) fixes the JES offset but further reduces the JES calibration
  slope. }
\begin{tabular}{lccc}
\hline \hline
        & JES slope & JES offset & $\delta$\mtop(*) \\
\hline
parton-matched \ttbar only & 0.96 & +0.026 & \\
\ttbar only & 0.88 & +0.050 & \\
all events & 0.80 & +0.076 & 4.30 GeV\\
\hline
all, 50\% offset correction  & 0.72 & +0.036 & 4.10 GeV\\
all, 100\% offset correction & 0.63 & +0.000 & 4.01 GeV \\
\hline \hline
\end{tabular}
(*) expected mass uncertainty after full calibration
\label{tab:JESslope}
\end{center}
\end{table}

The JES offset and slope turn out to be independent of the
generated top quark mass (see Fig.~\ref{fig:JesCal}).
Therefore we apply a straightforward mass-independent correction. A normalization factor $f_{\rm
  JES}(\jes, \ftop) = {\rm exp}(a \cdot \jes)$ is introduced which
corrects for the offset without changing
the statistical uncertainty estimated from the likelihood (in case
the final sample likelihood is Gaussian):
\begin{multline}
  \label{eq:id6}
 {\cal L}^{\rm corr}_{\rm evt}(m_{t},\jes,\ftop) = \\ f_{\rm JES}(\jes,\ftop) \cdot {\cal L}_{\rm
  evt}(m_{t},\jes,\ftop) . 
\end{multline}
Since background events on average cause a larger bias than signal events, $a$
is defined to be dependent on the measured signal fraction \ftop: $a =
  2.63 + 0.56(1-\ftop)$.
The value of the correction constant is tuned using MC simulation to give an
unbiased measurement of the JES at the reference scale $\jes$ = 1.
As shown in Table~\ref{tab:JESslope}, the application of this offset
correction removes the JES offset, but it further reduces the JES
calibration slope. Table~\ref{tab:JESslope} also shows that after full calibration
(described in the next section), the expected statistical uncertainty
on the top quark mass improves slightly when applying the
corrections. For illustrative purposes we also include a 50\% offset
correction in the table, where $0.5 \cdot a$ is used instead of $a$. 

The correction described above ensures that the fit is well-behaved
and that, for values of the JES parameter near 1, the fit results will stay
well within the range for which the ($\jes$, $m_t$) likelihood is
calculated. It does not, however, provide a full calibration of the analysis, which is described in Sec.~\ref{section:MC}.

\subsection{Combined likelihood fit}

Since each event is independent, the combined likelihood for the entire sample is calculated as the product of the single event likelihood curves:
\begin{equation}
  \label{eq:id5}
{\cal L}_{\rm samp}(m_{t},\jes,\ftop) = \prod_j  {\cal L}^{\rm corr}_{{\rm evt} j}(m_{t},\jes,\ftop) .
\end{equation}
This likelihood is maximized with respect to the top quark mass
$m_{t}$, the jet energy scale parameter $\jes$, and the estimated fraction of signal in the sample \ftop.

\subsection{Calibration using Monte Carlo simulation}
\label{section:MC}

The analysis is calibrated using Monte Carlo simulation.
Both the bias on the measured mass and the correctness of the estimated
statistical uncertainty are studied using ensemble tests, 
in which many simulated experiments (pseudoexperiments) are created,
each matching the size of the observed data sample.
%where each ensemble corresponds to a simulated experiment matching the
%size of the observed data sample.
Thousands of pseudoexperiments are constructed, combining \ttbar and \wjets
events from MC simulation. 
%(with Higgs tuning)
%and QCD events from the data skim with reversed lepton isolation
%cuts. 
The event fractions for \ttbar and \wjets 
%and QCD 
are allowed to fluctuate according to binomial statistics around the estimated fractions in
the actual data sample.
The fractions used are those listed in Table~\ref{tab:selection}.
 In the pseudoexperiments, the QCD multijets contribution is replaced by \wjets events.
 This deviation in QCD multijet fraction 
 is treated as a systematic uncertainty (see Sec.~\ref{sec:systuncs}).
The total sample size is fixed to the observed number of events in data (116 in electron+jets and 114 in
muon+jets).
To make optimal use of the  available MC statistics, standard resampling techniques are used, 
allowing for the multiple use of MC events when constructing the pseudoexperiments~\cite{Jackknife}. 
\begin{figure*}
  \includegraphics[width=5.9cm]{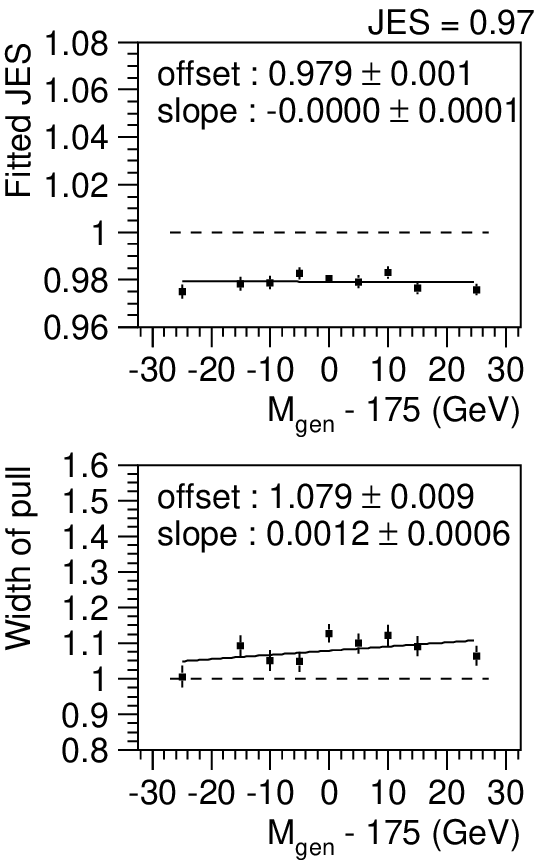}
  \includegraphics[width=5.9cm]{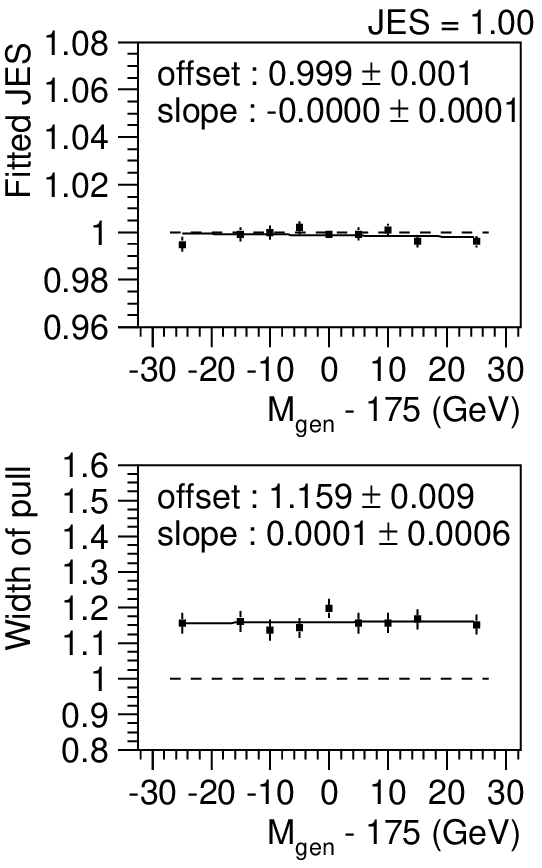}
  \includegraphics[width=5.9cm]{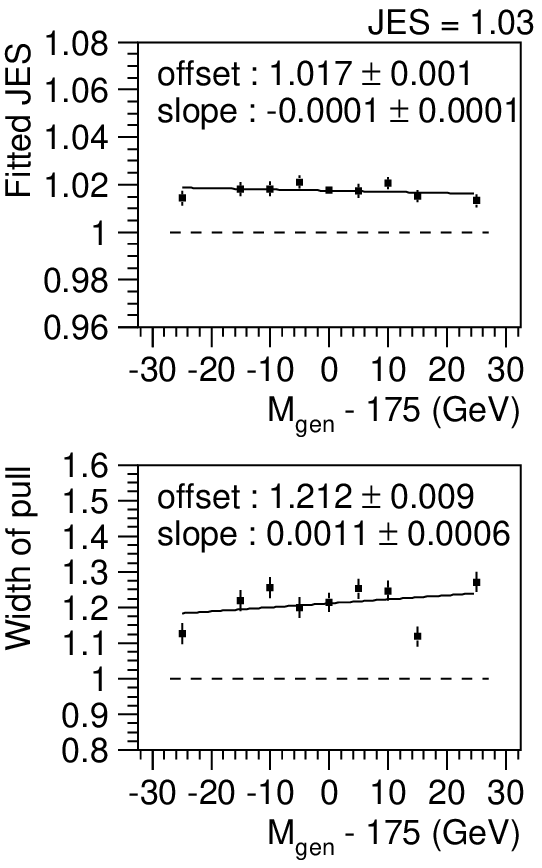}
\caption{\label{fig:JesCal} {The mean fitted $\jes$ and pull width as
    a function of the generated top quark mass $M_{\rm gen}$ for a ``true'' $\jes$ of 0.97
    (left), 1.00 (middle), and 1.03 (right), for the lepton+jets
    channel ($e$ + $\mu$ combined). 
	The fitted $\jes$ is stable as a function of generated
    top quark mass.}}
\end{figure*}
\begin{figure*}
  \includegraphics[width=5.9cm]{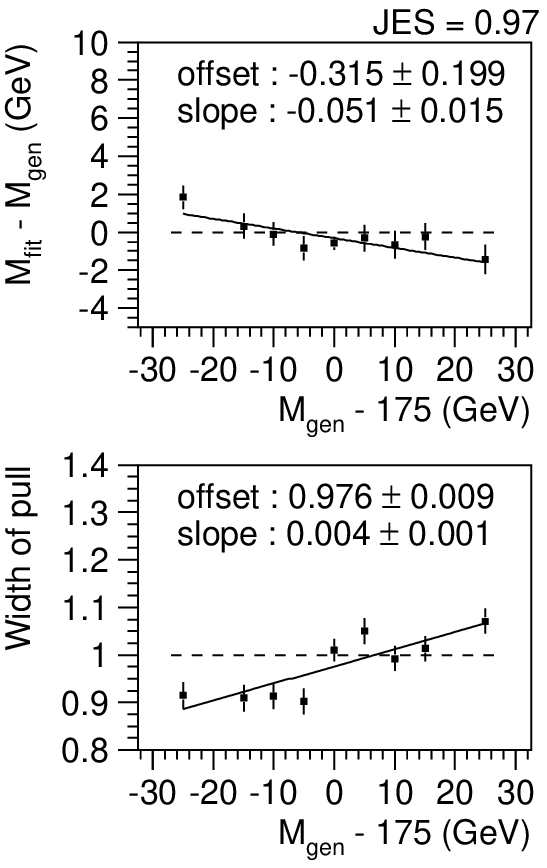} 
  \includegraphics[width=5.9cm]{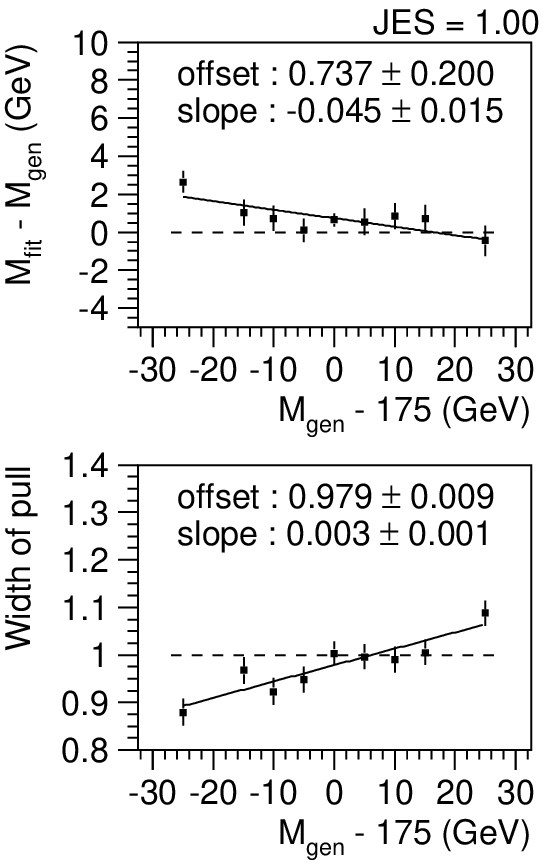}
  \includegraphics[width=5.9cm]{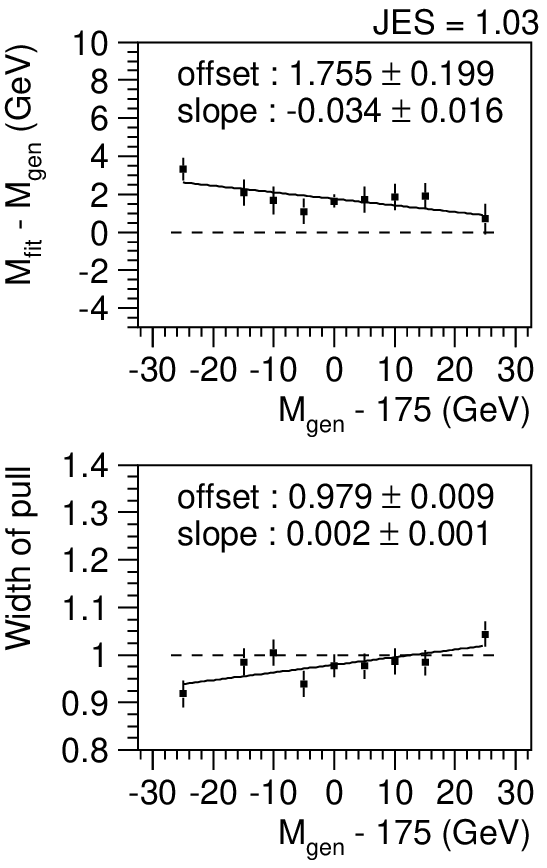}
\caption{\label{fig:MassCal} {The difference between the mean fitted mass
    $M_{\rm fit}$ and the generated top quark mass $M_{\rm gen}$ as
    a function of the generated top quark mass for a ``true'' $\jes$ of 0.97
    (left), 1.00 (middle), and 1.03 (right), for the lepton+jets
    channel ($e$ + $\mu$ combined).
At a generated mass of 175~GeV, the mass bias changes by 1 GeV when the true $\jes$ is varied by $\pm$3\%.}}
\end{figure*}
For every pseudoexperiment the mass is fitted and the deviation of this mass
from the mean of all pseudoexperiment masses is divided by the fitted
uncertainty. This quantity is referred to as the ``pull.''
The pull distribution for all pseudoexperiments is fitted with a Gaussian to extract the width, which we call the ``pull width.''
The corresponding pull and pull width for the fitted $\jes$ are also determined.

Figure~\ref{fig:JesCal} shows how the mean fitted $\jes$ and its pull
width behave as a function
of the top quark mass for different values of the true jet energy
scale. The fitted JES parameter is independent of the top quark mass over the full range
considered. The plots also show that the fitted $\jes$ changes 
linearly as a function of the true $\jes$ with a slope of 0.63 (see
discussion in Sec.~\ref{section:JESbias}).
Figure~\ref{fig:MassCal} shows the change in the fitted top
quark mass and the width of the pull as a function of the
generated top quark mass for different values of the true $\jes$. 
Using these plots
a full two-dimensional calibration is performed, describing the
fitted $\jes$ and top quark mass as a function of the ``true'' $\jes$ and top
quark mass generated in the MC simulation. The estimated statistical
uncertainties are corrected for the width of the pull and error
propagation is used to take into account the effect of the
two-dimensional calibration including the correlation between the JES
parameter and the offset in measured mass. 

\subsection{Alternative JES fitting strategies}
\label{section:alternative}

\begin{figure*}
  \includegraphics[width=5.9cm]{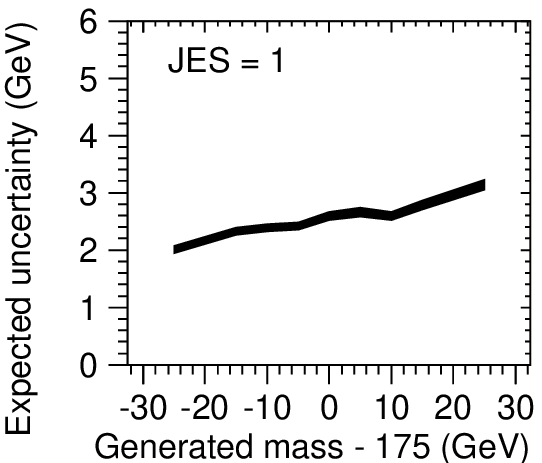}
  \includegraphics[width=5.9cm]{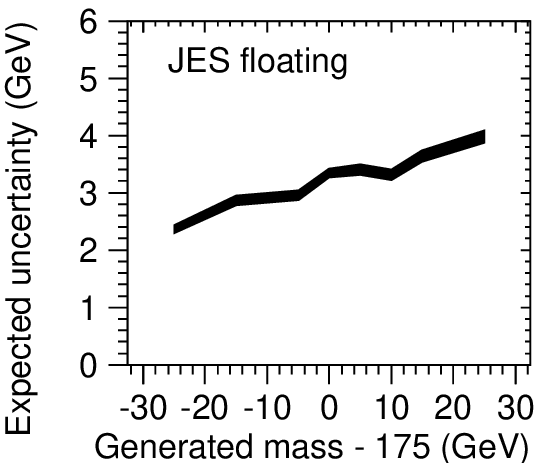}
  \includegraphics[width=5.9cm]{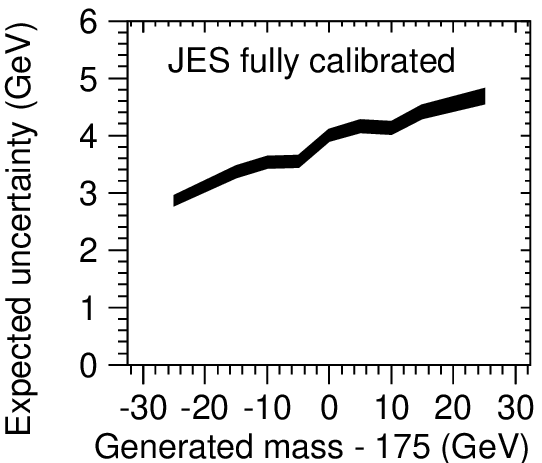}
\caption{\label{fig:Mass3Scenarios} {The  expected
    statistical uncertainty from ensemble tests is shown as
    a function of the generated top quark mass for three scenarios: with
    the JES parameter fixed to 1 (left), allowing the JES parameter
    to float freely in the fit, but only calibrating the mass fit for
    a true $\jes=1$ (middle), allowing the JES parameter to float
    freely in the fit and applying the full calibration as a function
    of true top quark mass and true $\jes$ (right). In each plot the
    width of the band indicates the estimated uncertainty. }}
\end{figure*}

Including the uniform JES parameter as a free parameter in the fit
reduces the systematic uncertainty due to the jet energy scale, at the
cost of a larger statistical uncertainty. As a comparison, in
Fig.~\ref{fig:Mass3Scenarios} the expected statistical uncertainties
on the top quark mass are shown for three different fitting
scenarios. When fixing the JES parameter in the fit to 1, the
statistical uncertainty is smallest: 2.59~GeV at a generated top quark mass
of 175~GeV. When allowing the JES parameter to
float freely in the fit, without correcting for the JES slope (=0.63) 
in the calibration, part of the sensitivity to the overall JES scale
is reduced and absorbed in the statistical uncertainty, leading to an
expected statistical uncertainty of 3.34~GeV at a top quark mass of
175~GeV. 
Fully calibrating the
analysis as a function of fitted mass and $\jes$ (the default approach), 
allows an unbiased top
quark mass measurement for any value of the ``true'' $\jes$, at the cost of a larger statistical uncertainty: 4.01~GeV at a generated top quark mass
of 175~GeV. 

In order to be consistent with the approach used by the
Matrix Element analysis~\cite{MEnew}, thus facilitating a combination
of results, and to
minimize the dependence on the external JES constraint from jet+photon
studies, the third scenario is presented here as the main analysis
result, applying the full calibration as a function of fitted top
quark mass and $\jes$. Results using the other two JES fitting strategies are quoted as a
cross-check in Sec.~\ref{section:Xcheck}. 

\section{RESULTS WITH DATA}
\label{sec:data}
\begin{figure*}
 \includegraphics[width=\linewidth]{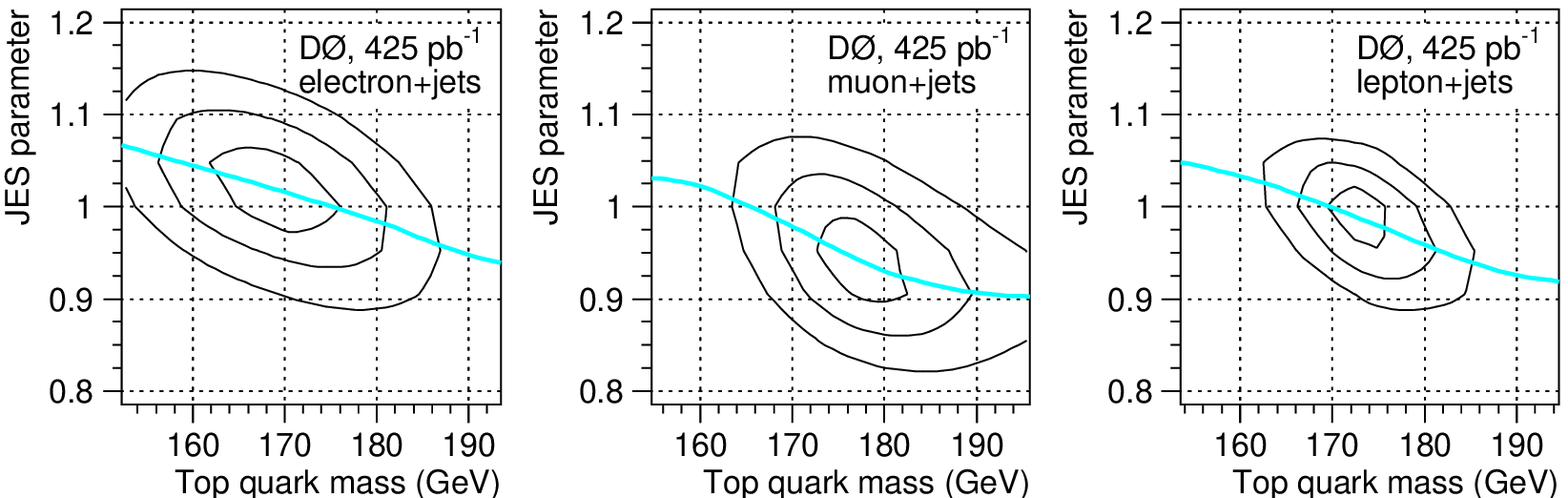}\\
% \vglue -.6cm
% \includegraphics[width=6.3cm]{figuresid/plot1djes.ejets.eps} \hglue -.7cm
% \includegraphics[width=6.3cm]{figuresid/plot1djes.mujets.eps} \hglue -.7cm
% \includegraphics[width=6.3cm]{figuresid/plot1djes.merged.eps}
 \includegraphics[width=\linewidth]{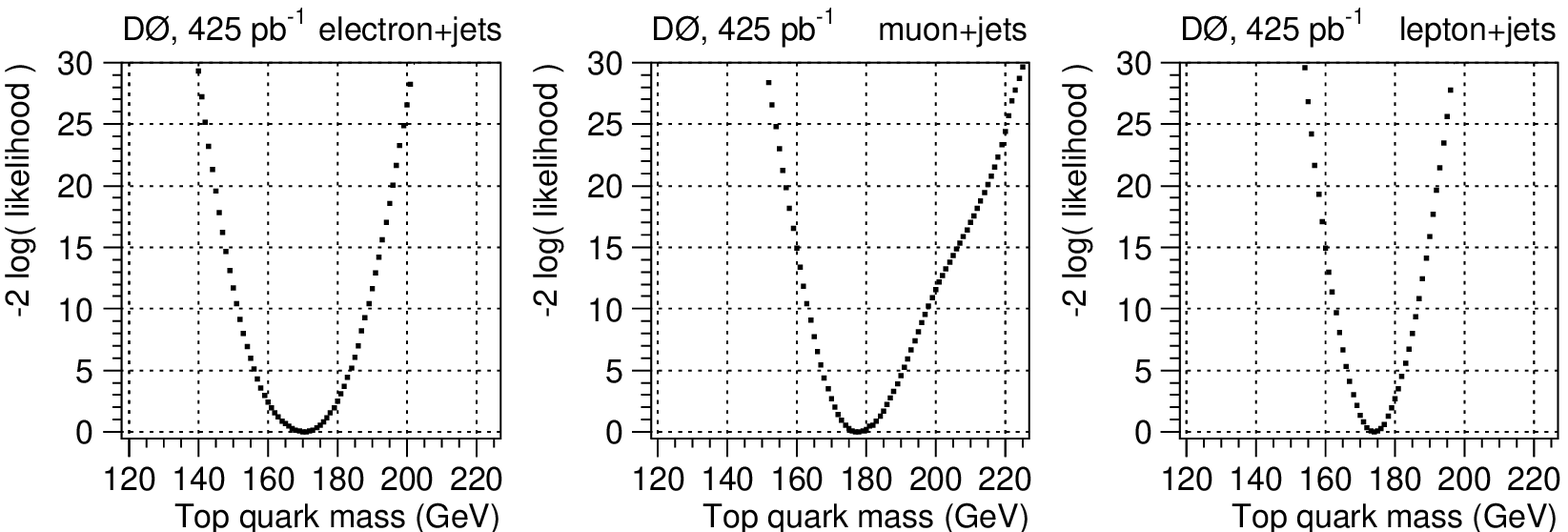}\\
\caption{\label{fig:LikEl} {Overall likelihood curves for
 the events observed in data, in the electron+jets channel (left), muon+jets
 (middle), and both channels combined (right).
The top plots show the full 2-dimensional likelihood as a function of the jet energy scale parameter ($\jes$) and top quark mass.
Each contour, $n$, corresponds to a difference in likelihood of
 $\Delta \ln({\cal L}) = -n^2/2$ with respect to the maximum likelihood.
%
%$\sqrt{2\cdot\text{ln(max~likelihood)}-2\cdot\text{ln(likelihood)}}$.
%
The fitted value of the JES parameter as a function of the top quark mass is plotted as the gray 
 line superimposed on the 2D likelihoods.
The bottom plots show the likelihood as a function of the top quark mass along the gray line from the upper plots.
The fitted values from these distributions have to be corrected for the calibration from MC simulation to obtain the final results. }}
\end{figure*}

The overall likelihood curves obtained for data are shown in
Fig.~\ref{fig:LikEl}.
\begin{figure*}
  \includegraphics[width=16cm]{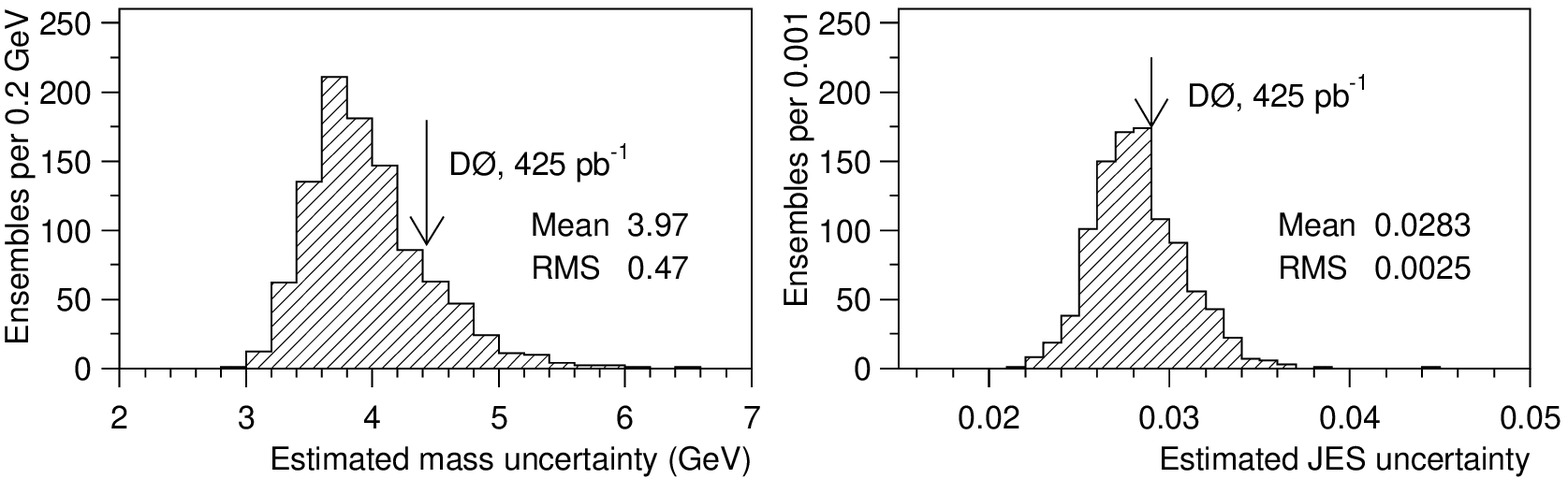}  
\caption{\label{fig:Error} {
  Distribution of the estimated statistical uncertainty on the top
quark  mass measurement (left) and JES measurement (right) for the fully
  calibrated analysis, in the combined lepton+jets
    channel. The values observed in data are indicated by the arrows.}}
\end{figure*}
The 2D likelihoods show the actual likelihood values in bins of 1~GeV in mass and 3\% in $\jes$.
The jagged appearance of the ellipses is caused by the large bin size 
in the $\jes$ direction.
To extract the mass and statistical error, a Gaussian
fit is applied to the three bins closest to the minimum in the
one-dimensional negative log likelihood curves. The fitted values are
corrected according to the calibration derived in Sec.~\ref{section:MC}. The measured top quark mass is: 
\begin{align}
m_{t} &= 173.7 \pm 4.4 \text{ (stat + JES) GeV} \notag \\
\text{with } \notag \\  \jes &=
0.989 \pm  0.029 \text{ (stat)}\notag .
\end{align}
All uncertainties shown are statistical. 
The fitted \ttbar signal fraction is $\ftop = 0.453 \pm 0.032$. If
the JES parameter is kept fixed to 1 in the fit, the estimated statistical
uncertainty is 2.93~GeV. Hence 
the 4.43~GeV (stat+JES) uncertainty of the 2D fit can be interpreted
as a combination of an intrinsic mass uncertainty of
2.93~GeV (stat) and an additional uncertainty of 3.32~GeV (JES) due to
fitting the JES parameter.
As shown in Fig.~\ref{fig:Error}, the observed statistical uncertainties are slightly larger than the average uncertainties expected from Monte Carlo ensemble tests, but they fall well within the distribution.
%%Only statistical uncertainties are shown here. 
The fitted $\jes$ of 0.989$\pm$0.029 is in good agreement with the reference
scale 1 (or ${\cal S}$=0), corresponding to the hypothesis that
after all jet corrections the JES in data and MC are the same. 

One can also compare the in-situ fitted JES parameter with the scale
obtained in jet+photon studies. When correcting all jets in MC events for the
jet-$p_T$ dependent difference between data and MC, ${\cal S}$, and redoing the ensemble tests in
MC simulation, the mean fitted $\jes$ is $0.962^{+0.021}_{-0.023}$, where the
uncertainties correspond to the combined statistical and systematic bounds from the
jet+photon studies. This is consistent with the value of
$0.989\pm 0.029$ measured in situ.

\section{Systematic Uncertainties}
\label{sec:systuncs}
The calibration of the analysis relies on Monte Carlo simulation. 
Therefore any discrepancy between the Monte Carlo simulation and
the data may lead to a bias and thus to a systematic shift in
the measured top quark mass.
In this section we describe the aspects of the simulation which may not accurately represent the data and evaluate the possible effect on the mass measurement.
To determine the impact of each uncertainty, we perform ensemble tests
 using a pool of simulated events that are modified according to the 
 uncertainty in question. The
 shift in the mean fitted ensemble mass compared to the default value 
 gives the size of the corresponding systematic uncertainty. 

The total systematic uncertainty on the top quark mass measurement is
obtained by adding all contributions in quadrature.
The following sources of systematics are considered (also see
%%the long summary in Table~\ref{tab:ideosystematics} and short 
%the summary in 
Table~\ref{tab:ideosystematicssummary}):

\begin{table*}
\begin{center}
\caption{Summary of systematic uncertainties.}
\begin{tabular}{lc}
\hline \hline
Source of uncertainty & Size of the
  effect (GeV) \\
%                                 & 1D Fit & 1.5D Fit & 2D Fit \\
                                                                                                
\hline
Jet energy scale ($p_T$ dependence)  & 0.45 
\\
%JES $p_T$ dependence (*)        & -              & -            & 0.45
%\\
Jet ID efficiency and resolution       & 0.22
\\
$b$ fragmentation                 & 1.30        
\\
$b$ response (h/e)                & 1.15 
\\
%$b/c$ semileptonic decays (**)    & $+0.06-0.07$
%\\
$b$ tagging                       & 0.29
\\
Trigger uncertainty             &$ +0.61 -0.28$  
\\
Signal modeling                 & 0.73       
\\
Signal fraction (stat+sys)      & 0.12
\\
Background modeling             & 0.20
\\
Multijet background            & 0.28
\\
MC calibration                  & 0.25
\\
PDF uncertainty                 & 0.023
\\
% \multicolumn{2}{|l|}{(**) taken from ME analysis}\\
%\hline
\\
Total systematic uncertainty  &$ +2.10 -2.04$ \\
%\hline
%Statistical uncertainty    &          \\ 
%including jet energy scale & 4.43  \\
%\hline
%Total uncertainty & +4.90 -4.87 \\
%{\bf residual JES} & {\bf +x.xx -x.xx} & {\bf +x.xx - x.xx} & {\bf
%+x.xx - x.xx}\\
\hline \hline
\end{tabular}
\label{tab:ideosystematicssummary}
\end{center}
\end{table*}
                                                                                                
%                                                                    
%\subsection{Physics Modeling}
%\label{sec:physicsmodelling}
%
\begin{itemize}

\item
  {\bf Jet energy scale {\boldmath $p_T$} dependence}: The inclusion of a
  uniform jet energy scale parameter $\jes$ as a free parameter in the
  mass fitting and calibration procedure ensures that a relative
  difference in overall jet energy scale between data and Monte Carlo is
  corrected for. The corresponding uncertainty 
  is included in the quoted statistical (stat + JES) uncertainty.
  Any residual discrepancy between data and Monte Carlo jet
  energy calibration that cannot be described by a uniform scale
  factor may lead to an additional systematic error on the top quark
  mass. The largest additional effect is expected from the uncertainty
  in the jet-$p_T$ dependence~\cite{MEnew}. The size of the impact of a  
  possible jet-$p_T$ dependent shape is estimated by scaling the energies of all 
  jets in the MC with a factor $(1 + 0.02 \frac{p^{\rm jet}_T - 100\
  {\rm GeV}}{100\ {\rm GeV}})$,
  where $p^{\rm jet}_T$ is the default reconstructed jet $p_T$. The value of 0.02 is suggested
  by the jet+photon studies. The mass obtained with the modified
  pseudoexperiments is compared to the default result and the shift of 
  0.45~\GeV is quoted as a systematic uncertainty.

\item
 {\bf Jet reconstruction efficiency and resolution:} In addition to
 uncertainties on the reconstructed jet energies, differences between
 data and the Monte Carlo simulation in the
 jet reconstruction efficiency and jet energy resolution may lead to a
 mass bias. Both efficiency and resolution are varied as a function of
 jet $p_T$ and rapidity within estimated uncertainties. No significant
 effect is observed, with an estimated statistical precision of
 0.15~\GeV. 
 For both effects combined, a systematic uncertainty of 0.22~\GeV is quoted.

\item
  {\bf {\boldmath $b$}-fragmentation:} 
  While the overall jet energy scale uncertainty is included in the statistical
  uncertainty from the fit, differences between data and Monte Carlo
  in the ratio of $b$-jet and light-jet energy scale could still affect the
  measurement. One possible source for such differences could be 
  the description of $b$-jet fragmentation in the
  simulation. To estimate the uncertainty from this source we used
  samples of simulated $\ttbar$ events with different fragmentation models
  for $b$ jets. The default Bowler~\cite{bib:BOWLER} scheme with
  $r_b$=1.0 is replaced with $r_b$=0.69 or with
  Peterson~\cite{bib:PETERSON} fragmentation with
  $\varepsilon_b$=0.00191. These parameter values were obtained by
  tuning \pythia simulation to LEP data~\cite{xA,xO,xD}. 
  The size of the variation in $r_b$
  corresponds to a larger shift in mean scaled energy $\langle x_B \rangle$
  of $b$ hadrons than the uncertainties reported in~\cite{xA,xO,xD,bib:Arno}. The comparison
  between the Bowler and Peterson scheme addresses the uncertainty on
  the shape of the $x_B$ distribution. Ensemble tests are repeated using events
  from each of the three simulations. The absolute values of the
  deviations in top quark mass results with respect to the standard sample
  are added in quadrature and quoted as a symmetric uncertainty of 1.3~\GeV.

%\item
%  {\bf {\boldmath$b/c$} semileptonic decays:}
%  The reconstructed energy of $b$ jets containing a semileptonic
%  bottom or charm decay is in general lower than that of jets 
%  containing only hadronic decays.  This can only be taken into account for
%  jets in which a soft muon is reconstructed.  Thus, the fitted top
%  quark mass still depends on the semileptonic $b$ and $c$ decay
%  branching ratios.  They are varied within the bounds given
%  in~\cite{bib:Zbible}, and the resulting shifts are found to be negligible
%  (evaluated using the Matrix Element method~\cite{MEnew}).

\item
  {\bf \boldmath{$b$}-jet energy response:} 
% h/e the average ratio between signals from electromagnetic and hadronic
% particles of the same incident energy
  Uncertainties in the simulation of the ratio between the calorimeter
  response to hadronic showers and electromagnetic showers (h/e ratio)
  may lead to additional differences in the $b$/light jet energy
  scale ratio between data and simulation. The possible size of the
  effect is studied in simulation, combining the uncertainty in the
  modeling of h/e calorimeter response ratio with the
  difference in particle content between light quark and $b$
  jets. An estimated uncertainty of 1.4\% on the $b$/light jet energy
  scale ratio is found. Ensemble tests show that this corresponds to a systematic
  uncertainty on the top quark mass of 1.15~\GeV.

\item
  {\bf\boldmath{$b$}-tagging:}
  The $b$ tagging rates for $b$ jets, $c$ jets, and light-quark
  jets are varied within the uncertainties known from the data, 
  and the resulting variations are propagated to the final mass
  results. Uncertainties in the heavy flavor composition of the
  background are also considered. 
  The combined effect is 0.29~\GeV.

\item   {\bf Trigger:} The trigger efficiencies in the Monte
  Carlo simulation are varied by their uncertainties estimated from
  data. The resulting variations in fitted mass are summed in
  quadrature, leading to a combined trigger uncertainty of
  $+0.61-0.28$~\GeV.

\item
  {\bf Signal modeling:} 
  The main uncertainty in the modeling of $\ttbar$ events is related
  to the radiation of gluons in the production
  or decay of the \ttbar system. A difference in the description of
  hard gluon radiation could affect the transverse momentum spectrum 
  of the \ttbar system or, for example, change the rate of confusion between jets 
  from the hadronically decaying $W$ boson and initial state gluons, 
  which could affect the  reconstructed top quark mass.
  To assess the uncertainty related to the modeling of high 
  energy gluons, the difference is studied between the default signal
  simulation and a dedicated $\ttbar$+jet simulation in which an
  energetic parton is produced in addition to the \ttbar system in the
  production process simulated by \alpgen. It is estimated that in the
  class of events that pass the full event selection, the fraction of
  simulated events with such an energetic gluon disagrees with the
  data by less than 35\%. 
%  The difference
%  in cross-section calculations between leading and next-to-leading
%  order suggests that such hard gluon radiation may be present in
%  about 35\% of \ttbar events that pass the event selection. 
  Pseudoexperiments are made with the usual sample
  composition, but replacing the default \ttbar events with the events
  from the dedicated $\ttbar$+jet simulation. 35\% of the observed
  shift in the fitted mass corresponds to 0.73~GeV, which is 
  assigned as a systematic uncertainty. 
\item
  {\bf Signal fraction:} 
  Since the \ttbar fraction $f_{\rm top}$ is fitted together with the top
  quark mass and the JES parameter, the mass measurement is
  affected by the uncertainty on the signal fraction in the data
  sample.
  We estimate two sources of systematic uncertainty: a variation of the signal fraction in the ensemble test used to calibrate the method
  and the effect of a possible systematic offset in the fitted signal fraction with
  respect to the true signal fraction internally in the mass fit. 

  We take the 7\% relative statistical uncertainty of the signal fraction
  found by combining the $\mu$+jets and $e$+jets numbers shown in Table~\ref{tab:selection}.
 We add in quadrature an estimated relative systematic uncertainty of
  11\% estimated from the cross section measurements~\cite{SVT}.
New ensemble tests for the calibration procedure, are performed with
  the mean of the Poisson distribution (for the signal fraction) 
  shifted by $(11 \oplus 7)\%$.
  Following this procedure the combined fit will still correctly fit
  the different signal fraction and compensate for the effect. This
  does not take into account the effect of a possible systematic
  discrepancy between the data and the Monte Carlo model of signal and
  background, which could lead to a systematic bias in the fitted 
  signal fraction.
%The systematic effect could also be underestimated by using the same Monte Carlo model in
%  the ensembles that is used to produce the $P_{\rm sgn}(D)$ and $P_{\rm bkg}$ Monte Carlo templates of
%  the in
%  the likelihood fit (see equations~\ref{eq:idsigfactorised} and~\ref{eq:idbgfactorised}).
To evaluate this additional systematic, the mass fit is forced to systematically 
  overestimate or underestimate the \ttbar fraction by 11\% 
  (with respect to the value preferred by the likelihood fit), and the 
  shift in fitted mass is quoted as a systematic
  uncertainty. The combined uncertainty, adding the above
  two contributions in quadrature, is 0.12~\GeV.
\item
  {\bf Background modeling:} 
  The sensitivity of the measurement to the choice of background model
  is studied by comparing two enlarged pseudoexperiments in which the background
  simulation is changed. One sample is based on the standard \wjets
  simulation using a factorization scale of $Q^2 = M_{\W}^{2} + \sum_{j}
  (p^j_T)^{2}$ while in the other pseudoexperiment a sample of \wjets events
  is used that are generated with a different factorization scale 
  of ${Q'}^2 = \langle p^j_T \rangle^{2}$. The observed difference in
  fitted mass is 0.20~\GeV which is assigned as a systematic
  uncertainty.
\item
  {\bf QCD multijet background:} In the calibration procedure, the 
  \wjets simulation is used to model the
  small multijet background in the selected data sample.
  To study the systematic uncertainty due to this approximation, we 
   selected a dedicated multijet-enriched sample of events from data by
  inverting the lepton isolation cut in the event selection. The
  calibration of the method is carried out with pseudoexperiments in which 
  these events are used to model the multijet background, according to the fractions
  given in Table~\ref{tab:selection}. The observed shift is 0.28~\GeV, which is quoted as a systematic uncertainty.
\item
  {\bf MC calibration:} The statistical uncertainty on the calibration
  curves shown in Fig.~\ref{fig:MassCal}
  is propagated through the
  analysis and yields a systematic uncertainty on the result
  of 0.25~\GeV.
\item
  {\bf Uncertainty due to the parton distribution functions (PDF):} The Ideogram analysis measures the top quark mass
  directly from the invariant mass of the \ttbar decay products
  without making specific assumptions regarding the production
  process. Nevertheless, the calibration of the analysis relies on Monte
  Carlo simulation in which a certain PDF set was used (CTEQ5L~\cite{bib:CTEQ5L}).
 It is  conceivable that a different choice of PDFs would lead to a slightly
  different calibration.  
  To study the systematic uncertainty on the top quark mass
  due to the precise PDF
  description, several PDF uncertainties are considered. PDF
  variations provided with the next-to-leading-order PDF set 
  CTEQ6M~\cite{bib:CTEQ6Mvar}
  are compared to the default CTEQ6M. The difference between CTEQ5L
  and MRST leading order PDFs is taken as a separate contribution.
  Also the effect of a variation 
  in $\alpha_s$ is evaluated. In all cases a large pseudoexperiment
  composed of events
  generated with CTEQ5L is reweighted so that distributions
  corresponding to the desired PDF set are obtained. The difference
  between weighted and unweighted pseudoexperiments is then quoted as
  systematic uncertainty, and all individual uncertainties are added
  in quadrature. The resulting combined uncertainty is found to be
  very small: $\pm 0.02$~\GeV.
\end{itemize}

\section{Cross-check using an external JES constraint}
\label{section:Xcheck}

As a cross-check, the analysis is repeated using the two alternative
JES fitting strategies discussed in Sec.~\ref{section:alternative}. 
Fixing the JES parameter in the fit and relying fully on the external
JES constraint from jet+photon studies, the top quark mass is measured
to be:
 $$ m_{\rm t} = 175.8 \pm 2.9 \mbox{ (stat)}^{+2.1}_{-2.7} \mbox{ (JES) GeV}, $$
%%\pm 2.9 \mbox{ (stat)}^{+2.9}_{-3.4} \mbox{ (syst) GeV}. $$
quoting only the statistical uncertainty (stat) and the systematic
uncertainty due to the jet energy scale (JES).
In the other alternative approach, the JES parameter is allowed to float freely 
in the fit but no calibration of the JES slope is applied.
Again, the external JES
constraint from jet+photon studies is required to set the jet energy scale and the remaining JES systematics.
Effectively this approach combines in-situ with external JES information, leading to the following result:
 $$ m_{\rm t} = 173.9 \pm 3.6 \mbox{ (stat)}^{+1.3}_{-1.0} \mbox{ (JES) GeV}. $$
%%\pm 3.6 \mbox{ (stat)}^{+2.2}_{-2.0} \mbox{ (syst) GeV}. $$
%In both cases the statistical uncertainty (stat) and the systematic
%uncertainty due to the jet energy scale (JES) are shown, but none of 
%the other systematic uncertainties are quoted.
%In both cases the systematic uncertainties include the increased JES
% systematic and the other systematic uncertainties, determined in the
% same way as described for the main result in
% Sec.~\ref{sec:systuncs}. 
 Comparing the last (most precise)
 cross-check with the main result, one can conclude that omitting the
 external JES constraint and relying fully on the in-situ
 information changes the central result only by 0.2 \GeV. The 2 \GeV
 difference between the first cross-check and the main result
 correlates very well with the 1.1\% difference in $\jes$ value between
 the default Monte Carlo scale and 
 in-situ JES measurements. This difference
 is fully covered by the quoted uncertainties.

\section{Conclusion}

\label{sec:conclusions}
The Ideogram method has been used for the first time to measure the top quark mass
in \ttbar events with the \ljets topology.
This technique employs a kinematic fit to extract mass information from the events,
while improving the statistical sensitivity by constructing an analytic
likelihood for every event taking into account all jet permutations and the
possibility that the event is background.  Lifetime-based
identification of $b$ jets is employed to enhance the separation between
\ttbar signal and background  and to improve the assignment of the observed jets to the partons in the \ttbar hypothesis.
To reduce the systematic uncertainty due to the jet energy scale
calibration, an overall scale factor $\jes$ for the energy of the
reconstructed jets is a free parameter in the fit 
determined simultaneously with the top quark
mass and the signal fraction. 

From a D0 Run II data sample of approximately 425~\ipb , 116 events are
selected in the electron+jets channel and 114 in the muon+jets channel.
The top quark mass is measured to be
 $$ m_t = 173.7 \pm 4.4 \mbox{ (stat + JES)}^{+2.1}_{-2.0}
 \mbox{ (syst)
  GeV} $$
with a fitted JES scaling factor:
 $$ JES = 0.989 \pm 0.029 \mbox{ (stat only)} ,$$
which is consistent with the reference jet energy scale (=1.0) and with the results
from the jet+photon calibration ($\approx 0.962^{+0.021}_{-0.023}$).
% and the JES factor
%fitted by the Matrix Element analysis on the same data set~\cite{MEnew}. 
The mass result is in good agreement with the Matrix Element
measurement using the same data set~\cite{MEnew} and with other recent
top quark mass measurements~\cite{massCDF,dilepton}.

% acknowledgement_paragraph_r2.tex                                 2/6/07
%
We thank the staffs at Fermilab and collaborating institutions, 
and acknowledge support from the 
DOE and NSF (USA);
CEA and CNRS/IN2P3 (France);
FASI, Rosatom and RFBR (Russia);
CAPES, CNPq, FAPERJ, FAPESP and FUNDUNESP (Brazil);
DAE and DST (India);
Colciencias (Colombia);
CONACyT (Mexico);
KRF and KOSEF (Korea);
CONICET and UBACyT (Argentina);
FOM (The Netherlands);
PPARC (United Kingdom);
MSMT (Czech Republic);
CRC Program, CFI, NSERC and WestGrid Project (Canada);
BMBF and DFG (Germany);
SFI (Ireland);
The Swedish Research Council (Sweden);
Research Corporation;
Alexander von Humboldt Foundation;
and the Marie Curie Program.


\begin{thebibliography}{99}
%%%%%%%%%%%%%%%%%%%%%%%%%%%%%%%%%%%%%%%%%%%%%%%%%%%%%%%%%%%%%%%%%%%%%%%%%%%%%%%%

% list_of_visitor_addresses_r2.tex                            2/6/07
%
%
\bibitem[*]{alton}
Visitor from Augustana College, Sioux Falls, SD, USA
\bibitem[\S]{podesta-lerma}
Visitor from ICN-UNAM, Mexico City, Mexico.
%\bibitem[\dag]{kozminski}
%Visitor from Lewis University, Romeoville, IL, USA
\bibitem[\ddag]{voutilainen}
Visitor from Helsinki Institute of Physics, Helsinki, Finland.
\bibitem[\#]{wenger}
Visitor from Universit{\"a}t Z{\"u}rich, Z{\"u}rich, Switzerland.
%
\vskip 0.25cm


\bibitem{discovery} F.~Abe \etal \ (CDF Collaboration), Phys.\ Rev.\ Lett.\ \textbf{74}, 2626 (1995); \\
 S.~Abachi \etal \ (D0 Collaboration), Phys.\ Rev.\ Lett.\ \textbf{74}, 2632 (1995).
%\bibitem{Ideo} \dzero CONF note 4574 (2004). 
%\\ http://www-d0.fnal.gov/Run2Physics/WWW/results/prelim/TOP/T08/T08.pdf 
%
\bibitem{EWWG} The ALEPH, DELPHI, L3, OPAL, and SLD Collaborations,
  the LEP Electroweak Working Group, SLD Electroweak Group, and SLD
  Heavy Flavour Group, Phys.\ Rept.\ \textbf{427}, 257 (2006).
%
\bibitem{run1templmass}
B.~Abbott \etal \ (D0 Collaboration), Phys.\ Rev.\ D \textbf{58}, 052001 (1998); \\
S.~Abachi \etal \ (D0 Collaboration), Phys.\ Rev.\ Lett.\ \textbf{79}, 1197 (1997).
%
\bibitem{massCDF} 
A.~Abulencia \etal \ (CDF Collaboration), Phys.\ Rev.\ D \textbf{73}, 032003 (2006);
A.~Abulencia \etal \ (CDF Collaboration), Phys.\ Rev.\ Lett.\ \textbf{96}, 022004 (2006).
%
\bibitem{ME} 
V.~M.~Abazov \etal \ (D0 Collaboration), Nature \textbf{429}, 638 (2004);\\
J.~Estrada, Ph.D.\ thesis, University of Rochester, 2001, http://lss.fnal.gov/archive/thesis/fermilab-thesis-2001-07.pdf.
%
\bibitem{otherLJprecise}
A.~Abulencia \etal \ (CDF Collaboration), Phys.\ Rev.\ D \textbf{73}, 092002 (2006).
%
\bibitem{MEnew} V.~M.~Abazov \etal \ (D0 Collaboration),  Phys.\ Rev.\
  D \textbf{74}, 092005 (2006);\\
  P.~Schieferdecker, Ph.D.\ thesis, LMU M\"unchen, 2005, http://www-d0.fnal.gov/results/publications\_talks/the\-sis/schieferdecker/thesis.pdf.
   
%FERMILAB-THESIS 2005-46.
%%
\bibitem{DELPHI} 
  P.~Abreu \etal \ (DELPHI Collaboration), Eur.\ Phys.\ J.\ C \textbf{2}, 581 (1998); \\ 
  P.~Abreu \etal \ (DELPHI Collaboration), Phys.\ Lett.\ B \textbf{462}, 410 (1999); \\
  P.~Abreu \etal \ (DELPHI Collaboration), Phys.\ Lett.\ B \textbf{511}, 159 (2001); \\
  M.~Mulders, Ph.D.\ thesis, FOM \& University of Amsterdam, 2001, http://delphiwww.cern.ch/~delphd/the\-sis/mulders/ThesisMulders.ps.gz.
%
\bibitem{PDG} W.-M.~Yao \etal \ (Particle Data Group), J.\ Phys.\ G \textbf{33}, 1 (2006).
%
\bibitem{CDFIdeogram} 
A.~Abulencia \etal \ (CDF Collaboration), submitted to Phys.\ Rev.\
Lett., hep-ex/0612026, FERMILAB-PUB-06/468-E.
%

% Detector
%
\bibitem{run2det} 
 V.~M.~Abazov \etal, Nucl.\ Instrum.\ Methods A {\bf 565}, 463 (2006).
\bibitem{run1det} 
 S.~Abachi \etal \ (D0 Collaboration), Nucl.\ Instrum.\ Methods A \textbf{338}, 185 (1994).
\bibitem{run2muon} V.~M.~Abazov \etal,  Nucl.\ Instrum.\ Meth.\ A \textbf{552}, 372 (2005).
%submitted to physics/0503151, FERMILAB-PUB-05/034-E.
%
\bibitem{kalman}
	R.~E.~Kalman, J.\ Bas.\ Eng.\ \textbf{82}, 35 (1960);\\
	R.~E.~Kalman and R.~S.~Brucy, J.\ Bas.\ Eng.\ \textbf{83}, 95 (1961);\\
	P.~Billoir, Nucl.\ Instrum.\ Meth.\ A \textbf{225}, 352 (1984).
\bibitem{btagxsecprd}
	V.~M.~Abazov \etal \ (D0 Collaboration),  Phys.\ Rev.\ D \textbf{74}, 112004 (2006).
% Rev.\ D,  hep-ex/0611002, FERMILAB-PUB-06/386-E.
%; \\
\bibitem{blazey} We use the iterative midpoint cone algorithm, as described in Sec. 3.5 of G.~C.~Blazey \etal,
in Proceedings of the Workshop: {\em QCD and Weak Boson Physics in Run II}, edited by U.~Baur, R.~K.~Ellis, and
D.~Zeppenfeld, Fermilab-Pub-00/297 (2000).
%
\bibitem{SVT}  V.~M.~Abazov \etal \ (D0 Collaboration), Phys.\ Lett.\ B \textbf{626}, 35 (2005).
\bibitem{ALPGEN} M.~L.~Mangano \etal, JHEP \textbf{307}, 1 (2003).
\bibitem{PYTHIA} T.~Sj\"{o}strand \etal, Computer Phys.\ Commun.\ \textbf{135}, 238 (2001).
\bibitem{geant} R.~Brun, F.~Carminati, CERN program library long writeup, W5013 (1993).
%
% Kinematic Fit
%
\bibitem{Whel}  V.~M.~Abazov \etal \ (D0 Collaboration), Phys.\ Rev.\ D \textbf{72}, 011104 (2005).
%
\bibitem{Charge} V.~M.~Abazov \etal \ (D0 Collaboration), to be published in Phys.\ Rev.\ Lett.,
  hep-ex/0608044, FERMILAB-PUB-06/278-E.  
%
\bibitem{topological} V.~M.~Abazov \etal \ (D0 Collaboration), Phys.\ Lett.\ B \textbf{626}, 45 (2005).
%
%
% Ideogram likelihood
%
%
%\bibitem{barlow} {\em SLUO Lectures on Statistics and Numerical
%  Methods in HEP, Lecture 6: Resampling and the Bootstrap}, Roger Barlow {\em and references
%  therein}, see http://www.hep.man.ac.uk/u/roger/home.html \\ and
%  http://www.hep.man.ac.uk/preprints/manhep99-4.ps
\bibitem{Jackknife}
B.\ Efron, {\em Computer and the Theory of Statistics}, SIAM Rev. {\bf
21}, 460 (1979);\\
P.\ Diaconis and B.\ Efron, {\em Computer-Intensive Methods in
Statistics}, Scientific American {\bf 248:5}, 96 (1983);\\
B.\ Efron and R.\ J.\ Tibshirani, {\em An Introduction to the
Bootstrap}, Chapman \& Hall/CRC, 1993.
%
%
%  Systematic Uncertainties
%
\bibitem{bib:BOWLER} M.G.~Bowler, Z.\ Phys C\textbf{11}, 169 (1981).
\bibitem{bib:PETERSON} C.~Peterson \etal, Phys.\ Rev.\ D \textbf{27}, 105 (1983).
%
\bibitem{xA}
A.~Heister \etal,
%``Study of the fragmentation of b quarks into B mesons at the Z
%peak,''
Phys.\ Lett.\ B \textbf{512}, 30 (2001).
%%[arXiv:hep-ex/0106051],
%%CITATION = HEP-EX 0106051;%%
\bibitem{xO}
G.~Abbiendi \etal \ (OPAL Collaboration),
%``Inclusive analysis of the b quark fragmentation function in Z
%decays at LEP.
%((B)),''
Eur.\ Phys.\ J.\ C \textbf{29}, 463 (2003).
%[arXiv:hep-ex/0210031],
%%CITATION = HEP-EX 0210031;%%
\bibitem{xD}
G.~Barker \etal, DELPHI conference report, 2002-69 CONF 603.
\bibitem{bib:Arno} 
  K.~Abe \etal \ (CDF Collaboration),  Phys.\ Rev.\ Lett.\ \textbf{84}, 4300 (2000).
%;\\
%  A.~Heister \etal,  Phys.\ Lett.\ B \textbf{512}, 30 (2001);\\
%  G.~Abbiendi \etal, Eur.\ Phys.\ J.\ C \textbf{29}, 463 (2003).
%
\bibitem{bib:CTEQ5L}
  H.~L.~Lai \etal \ (CTEQ Collaboration),
% ~[CTEQ Collaboration],
%  {\it Global {QCD} analysis of parton structure of the nucleon: CTEQ5
%  parton  distributions},
  Eur.\ Phys.\ J.\ C \textbf{12}, 375 (2000).
%
\bibitem{bib:CTEQ6Mvar} J.~Pumplin \etal \ (CTEQ Collaboration), JHEP 0207, 012 (2002).
%
\bibitem{dilepton}
A.~Abulencia \etal \ (CDF Collaboration), Phys.\ Rev.\ Lett.\ \textbf{96}, 152002 (2006);\\
A.~Abulencia \etal \ (CDF Collaboration), Phys.\ Rev.\ D \textbf{73}, 112006 (2006);\\
A.~Abulencia \etal \ (CDF Collaboration), Phys.\ Rev.\ D \textbf{74}, 032009 (2006);\\
A.~Abulencia \etal \ (CDF Collaboration), submitted to Phys.\ Rev.\
Lett., hep-ex/0612060, FERMILAB-PUB-06/490-E;\\
A.~Abulencia \etal \ (CDF Collaboration), submitted to Phys.\ Rev.\ D,
hep-ex/0612061, FERMILAB-PUB-06/487-E;\\
V.~M.~Abazov \etal \ (D0 Collaboration), submitted to Phys.\ Rev.\
Lett., hep-ex/0609056, FERMILAB-PUB-06/354-E.

  
%
%
%  OLD REFS
%
%
%\bibitem{mass} F.~Abe \etal, Phys.\ Rev.\ Lett.\ \textbf{80}, 2767 (1998); \\
%F.~Abe \etal, Phys.\ Rev.\ Lett.\ \textbf{80}, 2779 (1998);\\
%F.~Abe \etal, Phys.\ Rev.\ Lett.\ \textbf{82}, 271 (1999);\\
%F.~Affolder \etal, Phys.\ Rev.\ D \textbf{63}, 032003 (2001);\\
%S.~Abachi \etal, Phys.\ Rev.\ Lett.\ \textbf{79}, 1197 (1997);\\
%B.~Abbott \etal, Phys.\ Rev.\ Lett.\ \textbf{80}, 2063 (1998);\\
%B.~Abbott \etal, Phys.\ Rev.\ D \textbf{58}, 052001 (1998);\\
%B.~Abbott \etal, Phys.\ Rev.\ D \textbf{60}, 052001 (1999);\\
%V.~M.~Abazov \etal, Nature \textbf{429}, 638 (2004);\\
%V.~M.~Abazov \etal, Phys.\ Lett.\ B \textbf{606}, 25 (2005).
%
%\bibitem{discfun}
%Additional transformations to the $x_i$s before the fit were done for
%the functions to be better approximated by polynomials:
%$x_1'=exp[-(max(0,\sqrt{\frac{3(x_1/(1 \mathrm GeV)-5)}{2})]}$,
%$x_2'=\exp(-11x_2)$, $x_3' = \ln(x_3)$, $x_4'=\sqrt{x_4}$.
%\bibitem{jet+photon} conference note ?



%11] G. Mahlon, S. Parke, Phys. Lett. 80, 2063 (1997).
%[12] S. Eidelman \etal, Phys. Lett. B592, 1 (2004).
%[13] G.P. Lepage, Journal of Computational Physics 27:192-203, (1978).
%[14] G.P. Lepage, Cornell preprint CLNS:80-447, (1980).
%[15] H. L. Lai \etal, Eur. Phys. J. C 12, 375 (2000).
%\bibitem{VECBOS} F.A. Berends, H. Kuijf, B. Tausk, and W.T. Giele,
%Nucl. Phys. B357:32-64 (1991).
%\bibitem{MADGRAPH} F. Maltoni, T. Stelzer, JHEP 302, 27 (2003).

%\bibitem{bib:Zbible} The ALEPH, DELPHI, L3, OPAL, and SLD Collaborations, the LEP
%Electroweak Working Group, SLD Electroweak
%Group, and SLD Heavy Flavour Group, SLAC-R-774, (2005), hep-ex/0509008.


%%%%%%%%%%%%%%%%%%%%%%%%%%%%%%%%%%%%%%%%%%%%%%%%%%%%%%%%%%%%%%%%%%%%%%%%%%

\end{thebibliography}
\end{document}